\documentclass[12pt]{article}
\usepackage{eurosym}
\usepackage{graphicx}
\usepackage[utf8]{inputenc}
\usepackage[T1]{fontenc}
\usepackage{indentfirst}
\usepackage[margin=0.6in,nomarginpar]{geometry}
\usepackage[final]{hyperref}
\usepackage{amsmath}
\usepackage{hyperref}
\usepackage{cite}
\usepackage{subcaption}
\usepackage{amssymb}

\hypersetup{
colorlinks=true,
linkcolor=blue,
citecolor=blue,
filecolor=magenta,
urlcolor=blue
}
\begin{document}

\title{Thermodynamics and Shadows of quantum-corrected  Reissner-Nordström black hole surrounded by quintessence\ \ \thinspace }
\author{B. Hamil\thanks{%
hamilbilel@gmail.com} \\
Laboratoire de Physique Mathématique et Subatomique, Faculté des Sciences Exactes, \\
Université Constantine 1, Constantine, Algeria.  \and B. C. L\"{u}tf\"{u}o%
\u{g}lu\thanks{%
bekir.lutfuoglu@uhk.cz (Corresponding author)} \\
Department of Physics, University of Hradec Kr\'{a}lov\'{e},\\
Rokitansk\'{e}ho 62, 500 03 Hradec Kr\'{a}lov\'{e}, Czechia.}
\date{\today }
\maketitle

\begin{abstract}
In this work, we consider the quantum-corrected Reissner-Nordström black hole surrounded by quintessence matter and examine its thermodynamics and shadow. To this end, we handle the quantum-corrected metric of the black hole, which is given by Wu and Lui in 2022, and couple the quintessence matter field effects to the lapse function. Then, we obtain the mass, Hawking temperature, specific heat, and equation of state functions. In all these functions, we discuss two cases in which the quantum corrections are greater or smaller than the quintessence matter effects. We observe that within these two scenarios, the black hole has different characteristic thermal behaviors. Moreover, we find that the quintessence matter plays an important role in leading to a black hole remnant. { Furthermore, we observe that black hole behaves like Van der Waals fluid.} Finally, we investigate the black hole shadow by Carter's approach. We find that quintessence matter and quantum corrections modify the shadow of the black hole. 
\end{abstract}


\section{Introduction}
Earlier in the last century, Einstein proposed a new theory of gravity, known as the general theory of relativity, that overturned Newton's theory of universal gravity. In 1916, Schwarzschild obtained an exact solution to Einstein's equation which corresponds to a non-rotating black hole. In the same years, four independent scientists came up with another exact solution representing space-time for a charged, spherically symmetrical mass \cite{Reissner, Weyl, Nordstrom, Jeffery}. Later this solution, called Reissner-Nordström (R-N) black hole  solution,  is shown to permit stability and is found to allow supersymmetry within the context of $N=2$ supergravity \cite{Romans}. With the acquisition of other black hole metrics over time, the concept of black holes has become an interesting topic since they can be characterized by their mass, charge, and angular momentum. Although black holes cannot be observed directly with telescopes that detect X-rays, light, or other forms of electromagnetic radiation, one can infer the presence of black holes by detecting their effects on other matter nearby. In this context, in 2019 astronomers unveiled the first-ever image of the black hole at the center of the galaxy Messier 87 with the Event Horizon Telescope \cite{M871}. Last year, the image of the Sagittarius A*, a supermassive black hole located at the center of our Milky Way galaxy, was published \cite{SagA1}. Both images have similar properties:  shadows that are surrounded by  bright rings caused by gravity \cite{Chen2023}.

In the last half-century, black hole thermodynamics became another attraction center of modern cosmology and physics.  This field appeared after the revolutionary articles of Bekenstein and Hawking \cite{Bek, Bekenstein, Hawkink} in which the entropy and radiation of a black hole are defined proportionally by its event horizon area and surface gravity. Following these papers, many studies have been conducted to comprehend the laws of thermodynamics \cite{Hawking75, Aman2003, Dolan2011, Carlip2014, Mann2015, Kumar2020}. In particular, the quantum-corrected thermodynamics of Schwarzschild \cite{Adler2001, Medved2004, Myung2007, Nouicer2007, Ali2012, Moussa2021, EPL133, EPL134, EPL135}, RN \cite{Zhang2011, Gangopadhyay2014, Dutta2014, Lobo2021, EPJC2019, IJMPA36, EPJP136, PH1},   Kerr-Newman (K-N)\cite{Dehghani2011, Pourhassan2018, Ruiz, Wu2020} and  Ba\u{n}ados-Teitelboim-Zanelli (B-T-Z) {\cite{sarkar, akbar1, Eslam4, Eslam5, Iorio, Nadeem1, Nadeem2, Huang, BTZCan, BTZEUP, BTZEGUP, BP2, BP3} } black holes  have been investigated extensively in literature.

On the other hand,  reliable astronomical observations, that are performed a quarter of a century ago, have proven to us that the universe is expanding at an accelerating rate \cite{Riess1, Riess2, Perlmutter1999}. Some theoretical physicists have introduced the concept of dark energy to explain this phenomenon \cite{Peebles}. Basically, dark energy is a form of energy with negative pressure which is thought to constitute approximately sixty-eight percent of the universe. The origin of dark energy is first modeled with the cosmological constant,  however, the inconsistencies urged physicists to introduce new ideas such as modified matter, modified gravity, and the inhomogeneous models \cite{Yoo}. The quintessence matter model is a modified matter model, that is based on the dynamic scalar fields with a characteristic state parameter that indicates a ratio between the pressure and the energy density of the dark energy  \cite{carroll1998, Khoury2004, Picon2000, Padman2002, Caldwell2002, Gasperini2002, Copeland2006}. In 2003, Kiselev considered the quintessence matter model which surrounds the Schwarzschild black hole and investigated the black hole's thermodynamics \cite{Kiselev}. In the following years, in the presence of quintessence matter R-N \cite{Wei2011, Thomas}, K-N \cite{XuKerr}, B-T-Z \cite{Salwa}, Bardeen \cite{Ghaderi} and other \cite{Ghaderi16, Liu2019, Ndongmo, Zhang} black holes' thermodynamics are examined.

{

In classical general relativity, the problem of singularities, first highlighted by Penrose and Hawking in \cite{1,2}, always has been a matter of concern.  The main reason for these worries is that singularities lead to incomplete geodesics and hence to the disruption of causality. To cope with singularities, various types of corrections have been proposed for identifying regular, (i.e. nonsingular), black holes \cite{B, C, D, 3, 5, 6, 7,  9, Sudhaker, EPL23}. Many of regular black hole solutions can also be obtained as the solutions of the Einstein field equations coupled to suitable non-linear electrodynamics sources \cite{Stelle1, Stelle2, 10, Biswas1, 8}. In this manuscript, we consider the quantum correction, that was  given in Ref.\cite{10} by Kazakov and Solodukhin for examining the impact of quantum corrections on the behavior of the Schwarzschild black hole solution. In their work, they incorporated the spherically symmetric quantum fluctuations of the metric into the Einstein-Hilbert action and eliminated the point-like singularity. By doing so, they obtained the Kazakov-Solodukhin black hole, where instead of a central point-like singularity, there is a central 2D sphere with a radius of the order of the Planck length. This modification in the black hole's line element leads to a change in the event horizon geometry. Consequently, the thermodynamic properties of the black hole are altered. The introduction of this idea has drawn great attention in recent years, and many papers have appeared in the literature to investigate the
effects of the quantum corrections on black holes' physics and thermodynamics \cite%
{11,12,14,15,16,18,19,20,21,22,23,24,25,26,27}. Moreover, in the presence of the surrounding quintessence matter field quantum corrections to black hole physics and thermodynamics are also discussed in \cite{13, 13a, 13b, Shahjalal, EPJP136976, Shanping0, Shanping, Physlett, NuclPhysB, Li2023}.
}

Recent progresses show that the shadow of a black hole is a crucial phenomenological tool that helps to predict how the black hole could look in the presence of an illuminating source of light in its background \cite{He}. To our best knowledge spacetime properties and the observer's relative position to the black hole affect the shape and size of the black hole. For example, for a static observer the Schwarzschild black hole's shadow should be a perfect circle \cite{Synge}. However, for a rotating black hole, this expectation changes to a non-circular structure \cite{Bardeen}. After the detection of black holes, studies on black hole shadows gain importance and many researchers are working on  this issue, especially in the last five years {\cite{Singh1, Kogan1, Kogan2, Konoplya, WeiMann, Babar, Kumar, Kumar1, Singh, LiGuo, Zhang21, Thomas1,  Das, Volker, GuoWD, SS1, Heydar1, PG1, ZG1, Sunny1, Sunny2, Sunny3, Sunny4, Sunny5, Sunny6, Vir, Vir1}. }

Combining all these facts, we are motivated to investigate the shadow and thermal quantities of the quantum-corrected R-N black hole surrounded by quintessence matter. We present our study as follows: In the following section, we introduce the quantum-corrected R-N black hole and couple the quintessence matter field to its metric. Then, we derive the mass function to discuss the thermal quantities in section 3. After all, we investigate the black hole shadow in section 4. Finally, we present the conclusion.

\section{A brief review of quantum-corrected R-N black hole}
In four-dimensional space-time, the line element of the quantum-corrected R-N  black hole is found with the following form \cite{Shanping}%
\begin{equation}
ds^{2}=-f\left( r\right) dt^{2}+\frac{1}{f\left( r\right) }%
dr^{2}+r^{2}\left( d\theta ^{2}+\sin \theta d\varphi ^{2}\right) ,  \label{L}
\end{equation}
where the deformation is given in the lapse function by
\begin{equation}
f\left( r\right) =\frac{1}{r}\sqrt{r^{2}-a^{2}}-\frac{2M}{r}+\frac{Q^{2}}{r^{2}}.
\end{equation}
Here, $M$ and $Q$ denote the black hole's mass and charge, respectively,  while $a$ corresponds to the behavior of spherical symmetric quantum fluctuations via $a\equiv 4\ell _{p}$, where $\ell _{p}$ is the Planck length. We note that if one does not take into account the quantum fluctuations, then the usual lapse function is obtained. On the other hand, for small values of the quantum correction parameter, namely for $a<<r$, one can Taylor expand the deformed lapse function and arrive at the following form
\begin{equation}
f\left( r\right) \simeq \allowbreak 1-\frac{2M}{r}+\frac{Q^{2}}{r^{2}}-\frac{a^{2}}{2r^{2}},
\end{equation}%
which can be seen as the lapse function of an R-N black hole with an effective charge $q^{2}=Q^{2}-\frac{a^{2}}{2}$. 

\begin{figure}[tbh]
\begin{minipage}[t]{0.5\textwidth}
        \centering
        \includegraphics[width=\textwidth]{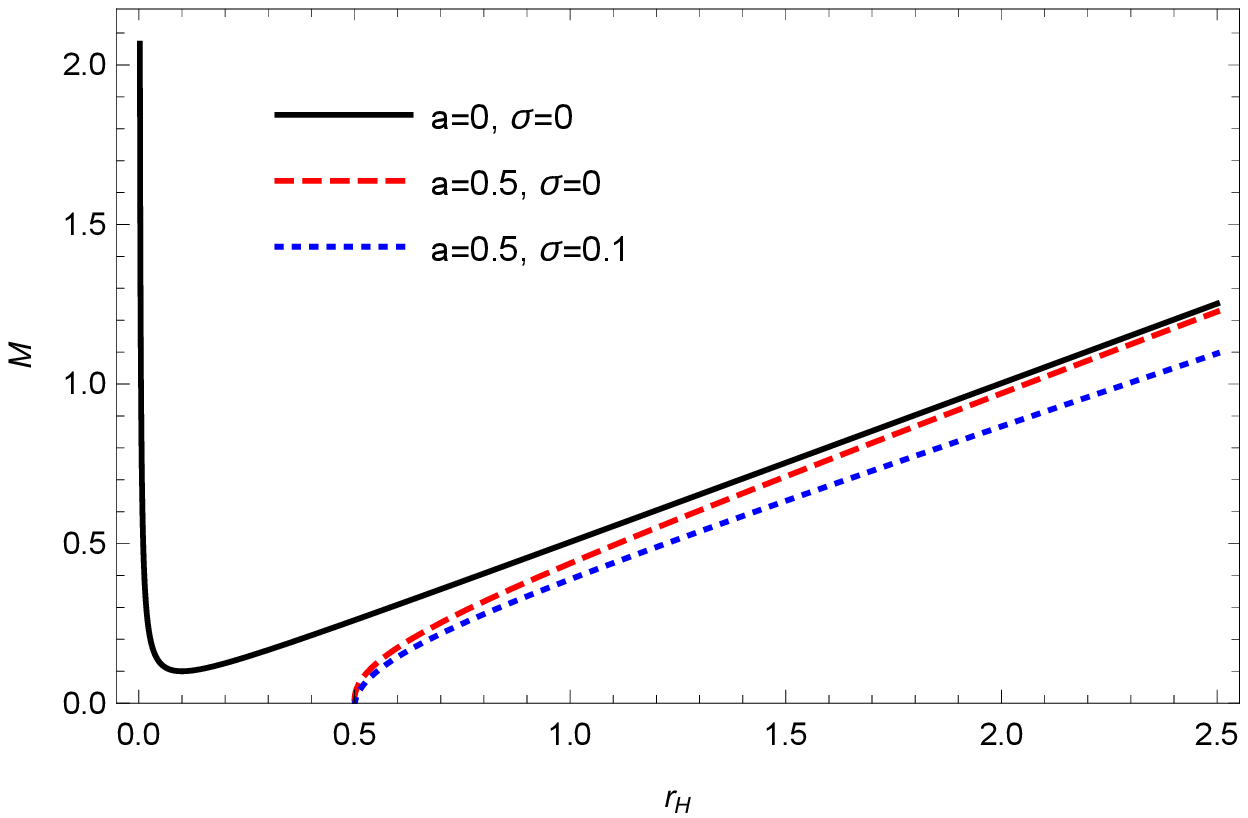}
       \subcaption{ $ \omega_q=-0.35$, and $Q=0.1$.}\label{fig:Ma}
   \end{minipage}%
\begin{minipage}[t]{0.50\textwidth}
        \centering
       \includegraphics[width=\textwidth]{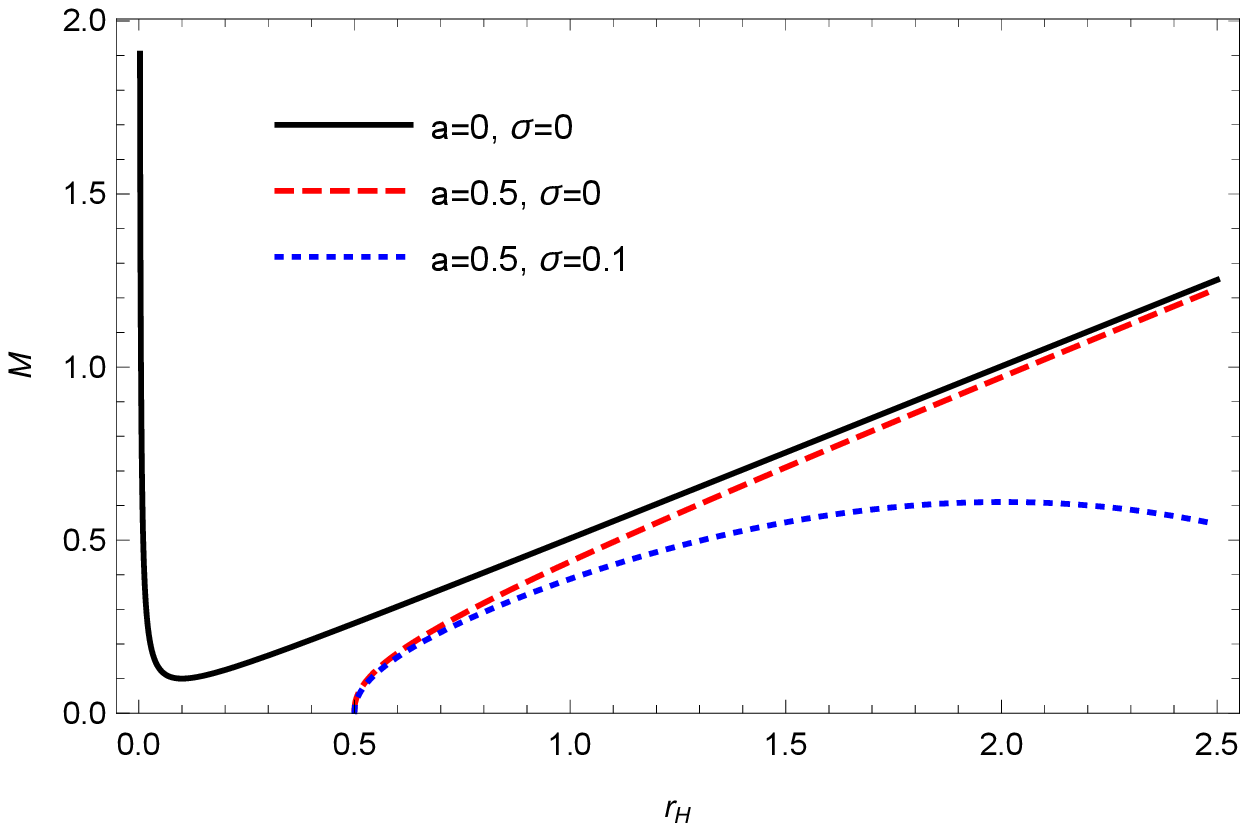}\\
        \subcaption{  $ \omega_q=-0.95$, and $Q=0.1$.}\label{fig:Mb}
    \end{minipage}\hfill
\begin{minipage}[b]{0.5\textwidth}
        \centering
        \includegraphics[width=\textwidth]{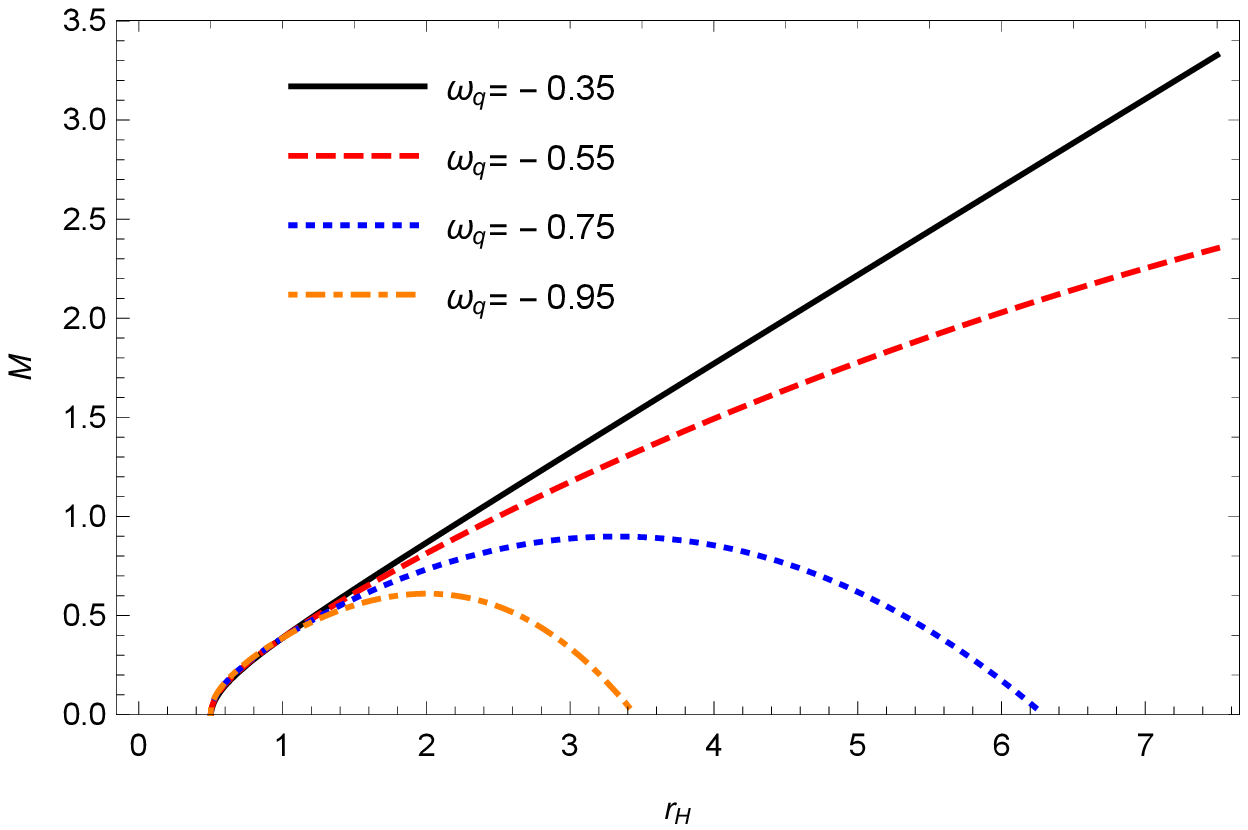}
       \subcaption{ $Q=0.1$, $ a=0.5$  and $\sigma=0.1$.}\label{fig:Mc}
   \end{minipage}%
\begin{minipage}[b]{0.50\textwidth}
        \centering
       \includegraphics[width=\textwidth]{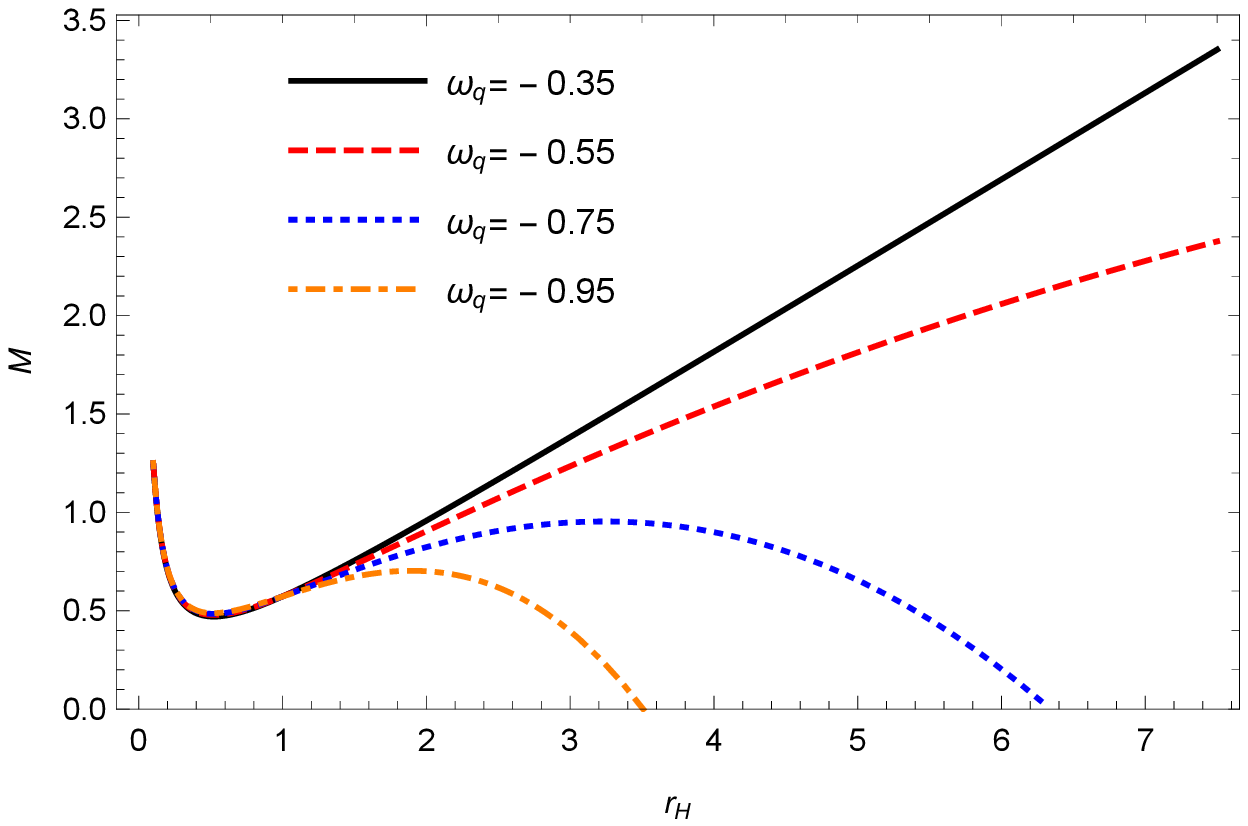}\\
        \subcaption{ $Q=0.5$, $ a=0.1$    and $\sigma=0.1$}\label{fig:Md}
    \end{minipage}\hfill    
\caption{Quantum-corrected R-N black hole mass function versus event horizon radius in the presence of quintessence matter.}
\label{Mfigsnew}
\end{figure}

Now, we consider the quantum-corrected  R-N black hole surrounded by the quintessence matter.  To express the geometry of the line element, we rewrite the lapse function in the following form  \cite{Shahjalal}:
\begin{equation}
f\left( r\right) =\frac{1}{r}\sqrt{r^{2}-a^{2}}-\frac{2M}{r}+\frac{Q^{2}}{%
r^{2}}-\frac{\sigma }{r^{3\omega _{q}+1}}.  \label{met}
\end{equation}
Here, $\sigma $ is a positive normalization factor that depends on the density of quintessence matter, while $\omega _{q}$ is the quintessential state parameter which takes values in the range of $\left( -1,-1/3\right) $. By setting $f\left( r_{H}\right) =0$, one can determine the event horizon radius, $r_{H}$, for a specific quintessential state parameter value. Then, the quantum-corrected mass of the black hole can be written in terms of the horizon as follows:
\begin{equation}
M=\frac{1}{2}\Bigg[\sqrt{r_{H}^{2}-a^{2}}+\frac{Q^{2}}{r_{H}}-\frac{\sigma }{%
r_{H}^{3\omega _{q}}}\Bigg].  \label{MASS}
\end{equation}
In Fig. \ref{Mfigsnew}, we plot the mass function versus horizon with different parameter values. 

In particular, Fig.~\ref{fig:Ma} and  Fig.~\ref{fig:Mb} show how the quantum corrections and quintessence matter affect the mass function. We see that the quantum-corrected mass function takes real values just after a minimal horizon value in the presence and absence of the quintessence matter. This means that quantum corrections are responsible to establish a minimal horizon value. On the other hand, we see that quintessence matter is responsible to settle an upper limit on the event horizon radius. We also observe that a greater valued  quintessential state parameter leads to a greater change in the behavior of the mass function at the greater horizon radius. In Fig.~\ref{fig:Mc} and  Fig.~\ref{fig:Md} we consider $Q<a$ and $Q>a$ cases, respectively. We see that in two cases the mass functions' characters do not differ from each other at small radius values.

Before we derive the thermal quantities, we would like to present the critical horizon radii and mass values of Fig.~\ref{fig:Mc} and Fig.~\ref{fig:Md} cases in Table \ref{Tabfig1c} and Table \ref{Tabfig1d}, respectively. Here, $r_{turn}$ stands for the turning point that corresponds to an important physical quantity which will be presented in the following section.
\begin{table}[tbp]
\center
\begin{tabular}{|c|c|c|c|c|c|c|}
\hline
$\omega_{q}$ & $r_{min}$ & $M(r_{min})$  & $r_{turn}$ & $M({r_{turn}})$  & $r_{max}$ & $M({r_{max}})$   \\ \hline
$-0.35$ & $0.5$ & $0$ & $ 10^{19}$ & $ 10^{18}$ & $ 10^{20}$ & $0$  \\ \hline
$-0.55$ & $0.5$ & $0$ & $16.002$ & $3.146$ & $34.546$  & $0$  \\ \hline
$-0.75$ & $0.5$ & $0$ & $3.326$  & $0.899$ & $6.295$   & $0$  \\ \hline
$-0.95$ & $0.5$ & $0$ & $2.003$  & $0.610$ & $3.453$   & $0$  \\ \hline
\end{tabular}
\caption{$Q=0.1$ and $a=0.5$ case.} \label{Tabfig1c}
\end{table}

\begin{table}[tbp]
\center
\begin{tabular}{|c|c|c|c|c|c|c|}
\hline
$\omega_{q}$ & $r_{min}$ & $M(r_{min})$  & $r_{turn}$ & $M({r_{turn}})$  & $r_{max}$ & $M({r_{max}})$   \\ \hline
$-0.35$ & $0.1$ & $1.246$ & $0.522$ & $0.470$ & $ 10^{20}$ & $0$  \\ \hline
$-0.55$ & $0.1$ & $1.249$ & $0.524$ & $0.479$  & $34.562$  & $0$  \\ \hline
$-0.75$ & $0.1$ & $1.250$ & $0.521$ & $0.484$   & $6.340$   & $0$  \\ \hline
$-0.95$ & $0.1$ & $1.250$ & $0.517$ & $0.488$   & $3.509$   & $0$  \\ \hline
\end{tabular}
\caption{$Q=0.5$ and $a=0.1$ case.} \label{Tabfig1d}
\end{table}

\section{Thermodynamics of quantum-corrected R-N black hole surrounded by quintessence}

\begin{figure}[tbh]
\begin{minipage}[t]{0.5\textwidth}
        \centering
        \includegraphics[width=\textwidth]{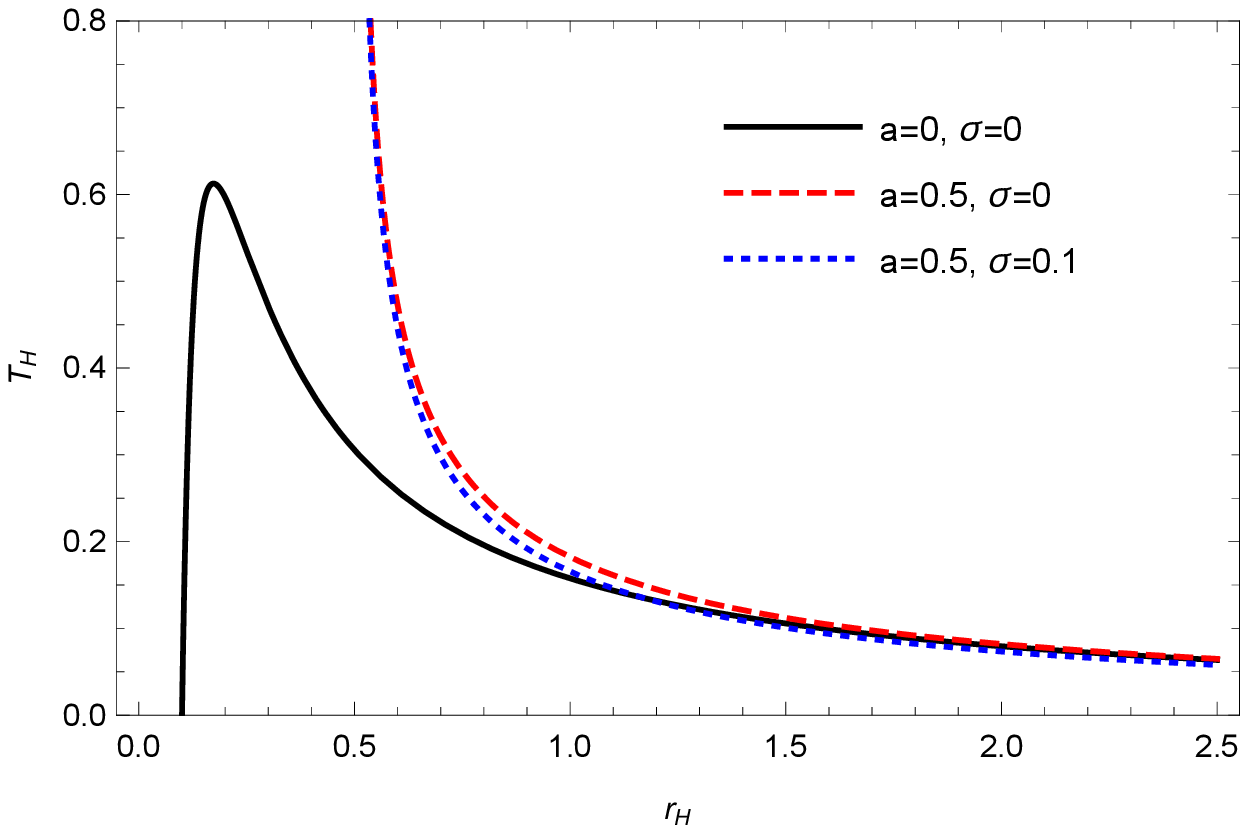}
       \subcaption{ $ \omega_q=-0.35$, and $Q=0.1$.}\label{fig:Ta}
   \end{minipage}%
\begin{minipage}[t]{0.50\textwidth}
        \centering
       \includegraphics[width=\textwidth]{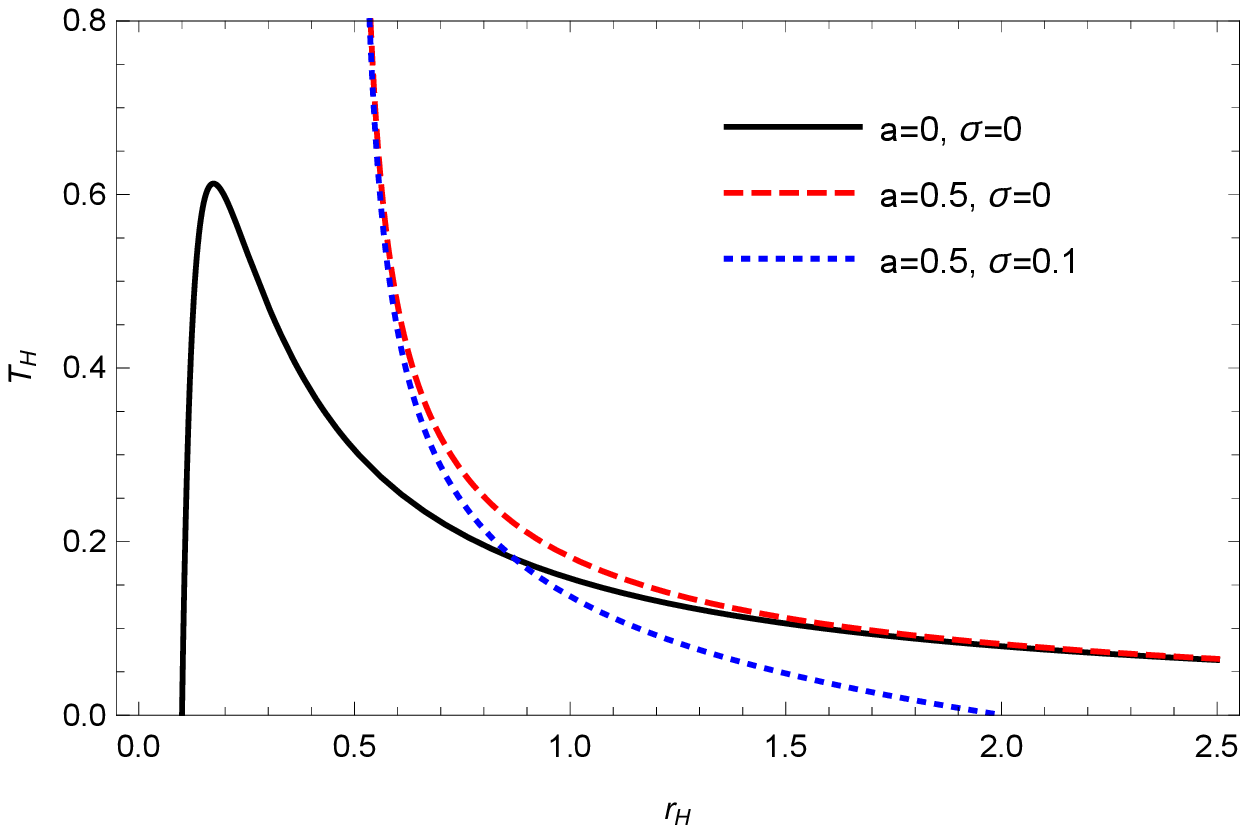}\\
        \subcaption{  $ \omega_q=-0.95$, and $Q=0.1$.}\label{fig:Tb}
    \end{minipage}\hfill
\begin{minipage}[t]{0.5\textwidth}
        \centering
        \includegraphics[width=\textwidth]{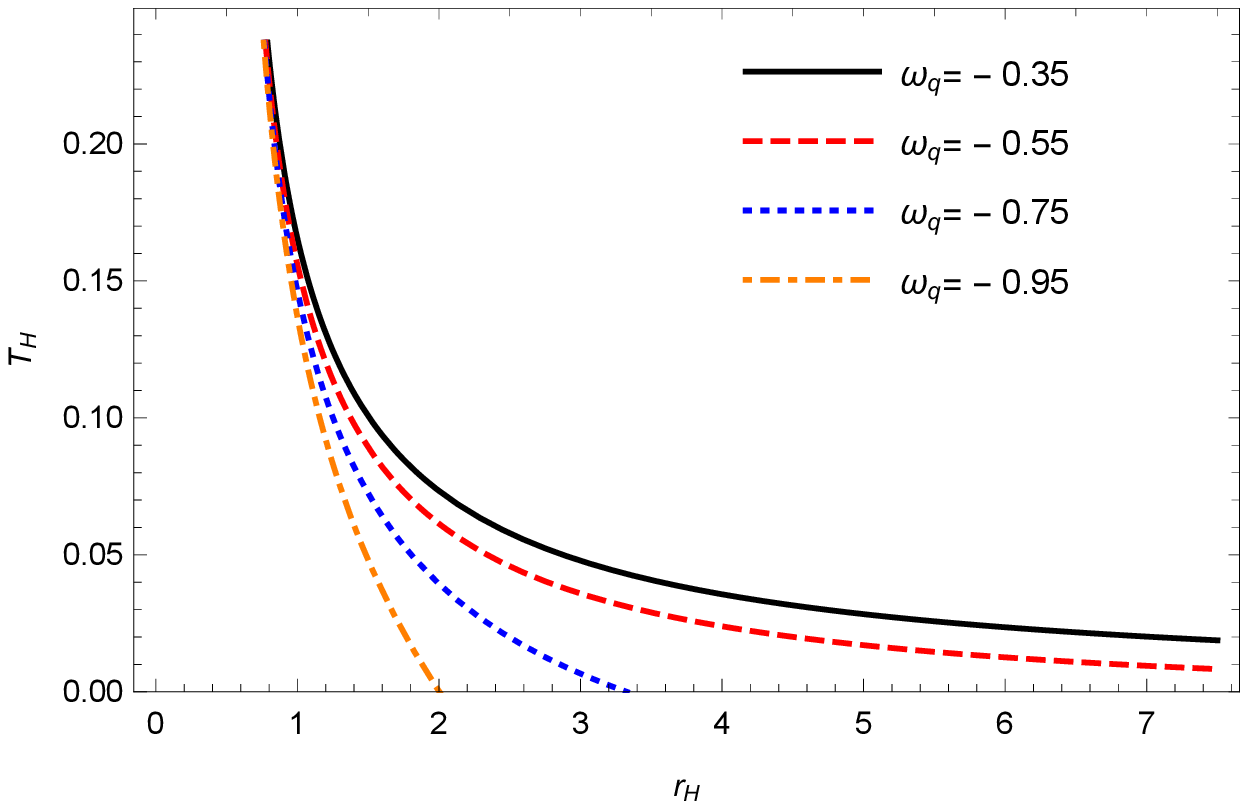}
       \subcaption{ $ a=0.5$, $Q=0.1$   and $\sigma=0.1$.}\label{fig:Tc}
   \end{minipage}%
\begin{minipage}[t]{0.50\textwidth}
        \centering
       \includegraphics[width=\textwidth]{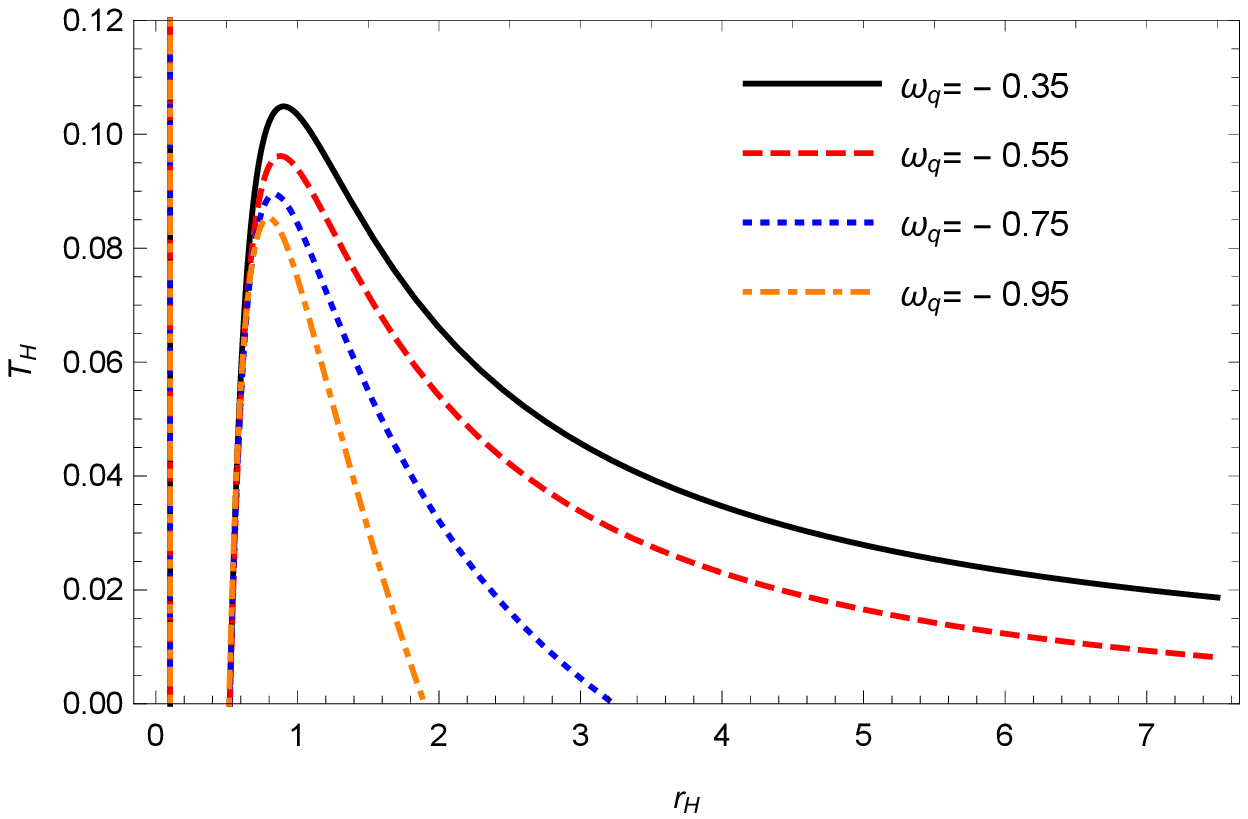}\\
        \subcaption{ $ a=0.1$, $Q=0.5$   and $\sigma=0.1$}\label{fig:Td}
    \end{minipage}\hfill    
\caption{Quantum-corrected Hawking temperature versus event horizon radius in the presence of quintessence matter.}
\label{Tfigsnew}
\end{figure}

Now, we intend to investigate the thermodynamics of the quantum-corrected R-N black hole surrounded by quintessence. We start with the Hawking temperature which can is defined as follows:
\begin{equation}
T=\frac{\kappa }{2\pi },  \label{t}
\end{equation}%
in the semiclassical framework. Here,  $\kappa $ denotes the surface gravity and can be calculated by the metric elements 
\begin{equation}
\kappa =-\lim_{r\rightarrow r_{H}}\sqrt{-\frac{g^{11}}{g^{00}}}\frac{\left(
g^{00}\right) ^{\prime }}{g^{00}}.
\end{equation}%
Performing the straightforward calculations, we arrive at the quantum-corrected Hawking temperature of the R-N black hole in the form of
\begin{equation}
T_{H}=\frac{1}{2\pi r_{H}}\left( \frac{a^{2}}{r_{H}^{2}\sqrt{1-\frac{a^{2}}{%
r_{H}^{2}}}}+\sqrt{1-\frac{a^{2}}{r_{H}^{2}}}-\frac{Q^{2}}{r_{H}^{2}}+\frac{%
3\sigma \omega _{q}}{r_{H}^{3\omega _{q}+1}}\right) .  \label{Sur}
\end{equation}%
We note that in the absence of the quantum-correction parameter and quintessence
field, Eq.(\ref{Sur}) reduces to the ordinary R-N black hole temperature. Before examining the next thermal quantity, we present the plots of quantum-corrected Hawking temperature versus horizon in Fig. \ref{Tfigsnew}. In particular, we demonstrate the influence of quantum correction and quintessence matter on the Hawking temperature for two quintessential state parameters in Fig. \ref{fig:Ta} and Fig. \ref{fig:Tb}. The second row is devoted to showing the cases $Q<a$ and $Q>a$, which have two distinct characteristic changes. For the $Q<a$ case, the Hawking temperature receives positive real values between two horizon radii which we indicate with $r_{asy}$ and $r_{rem}$ in Table \ref{Tabfig2c}. { We observe that the quantum-corrected Hawking temperature is a monotonically decreasing function in this interval. Moreover, we see that it  diverges in the limit of $r_H\rightarrow a$. Therefore, we prefer to name this root} as the asymptotic horizon radius. On the other hand, when we compare the greater root of the mass and the Hawking temperature functions, namely $r_{max}$ and $r_{rem}$, we observe that $r_{rem}<r_{max}$ relation holds. For the $Q>a$ case, $r_H=r_{min}$ is just a discontinuity. The physically meaningful temperature is found in the interval of $r_{phys}< r_H < r_{rem}$, where $r_{rem}$ is always smaller than $r_{max}$ as in the previous case. { We observe that the quintessence
parameter has an important role in the maximum value of the quantum-corrected Hawking temperature. More precisely, a greater quintessence parameter leads to a greater maximal temperature value.} We notice that, unlike the former case, the quantum-corrected Hawking temperature { is not a monotonic function, first it increases until a critical horizon radius, and then, it} decreases to zero. { One can determine the critical radius by
\begin{equation}
\left. \frac{\partial T_{H}}{\partial r_{H}}\right\vert _{r_{H}=r_{cr }}=0.
\end{equation}%
It should be mentioned that when a black hole temperature reaches its maximum value, the specific heat diverges. Therefore, the critical horizon value }  is responsible for a phase change and will be investigated via the heat capacity function in the context of the stability of the black hole. In Table \ref{Tabfig2d}, we present the horizons radii, the Hawking temperatures, and mass values for the considered four quintessential state parameters. In both cases, we observe that the Hawking temperature takes smaller values for smaller values of quintessence state parameters, especially at greater horizon radii. 

\begin{table}[tbp]
\center
\begin{tabular}{|c|c|c|c|c|c|}
\hline
$\omega_{q}$ & $r_{asy}$ & $T_{asy}$ & $r_{rem}$ & $T_{rem}$ & $M_{rem}$      \\ \hline
$-0.35$ & $0.5$ & $\infty$ & $ 10^{19}$  & $0$  & $10^{18}$ \\ \hline
$-0.55$ & $0.5$ & $\infty$ & $16.002$    & $0$  & $3.147$   \\ \hline
$-0.75$ & $0.5$ & $\infty$ & $3.326$     & $0$  & $0.899$   \\ \hline
$-0.95$ & $0.5$ & $\infty$ & $2.003$     & $0$  & $0.610$   \\ \hline
\end{tabular}
\caption{$Q=0.1$ and $a=0.5$ case.} \label{Tabfig2c}
\end{table}

\begin{table}[tbp]
\center
\begin{tabular}{|c|c|c|c|c|c|c|c|c|}
\hline
$\omega_{q}$ & $r_{asy}$ & $r_{phy}$  & $r_{cr}$ & $T_{cr}$ & $M_{cr}$ & $r_{rem}$ & $T_{rem}$ & $M_{rem}$   \\ \hline
$-0.35$ & $0.1$ & $0.522$ & $0.903$ & $0.105$ & $0.542$  & $10^{19}$  & $0$ &  $10^{18}$  \\ \hline
$-0.55$ & $0.1$ & $0.524$ & $0.881$ & $0.096$ & $0.539$  & $15.967$   & $0$ &  $3.157$  \\ \hline
$-0.75$ & $0.1$ & $0.521$ & $0.838$ & $0.089$ & $0.532$  & $3.236$    & $0$ &  $0.954$ \\ \hline
$-0.95$ & $0.1$ & $0.517$ & $0.796$ & $0.085$ & $0.526$  & $1.897$    & $0$ &  $0.703$ \\ \hline
\end{tabular}
\caption{$Q=0.5$ and $a=0.1$ case.} \label{Tabfig2d}
\end{table}
Then, we handle the entropy function by using the Bekenstein-Hawking formula \cite{Bekenstein}.
\begin{equation}
S=\pi r_{H}^{2}.  \label{S}
\end{equation}%
Here, we cannot provide an explicit form for the entropy function, since the event horizon radius cannot be expressed analytically. In other words, the entropy can be calculated only after determining particular values of the parameters. 

\begin{figure}[tbh]
\begin{minipage}[t]{0.5\textwidth}
        \centering
        \includegraphics[width=\textwidth]{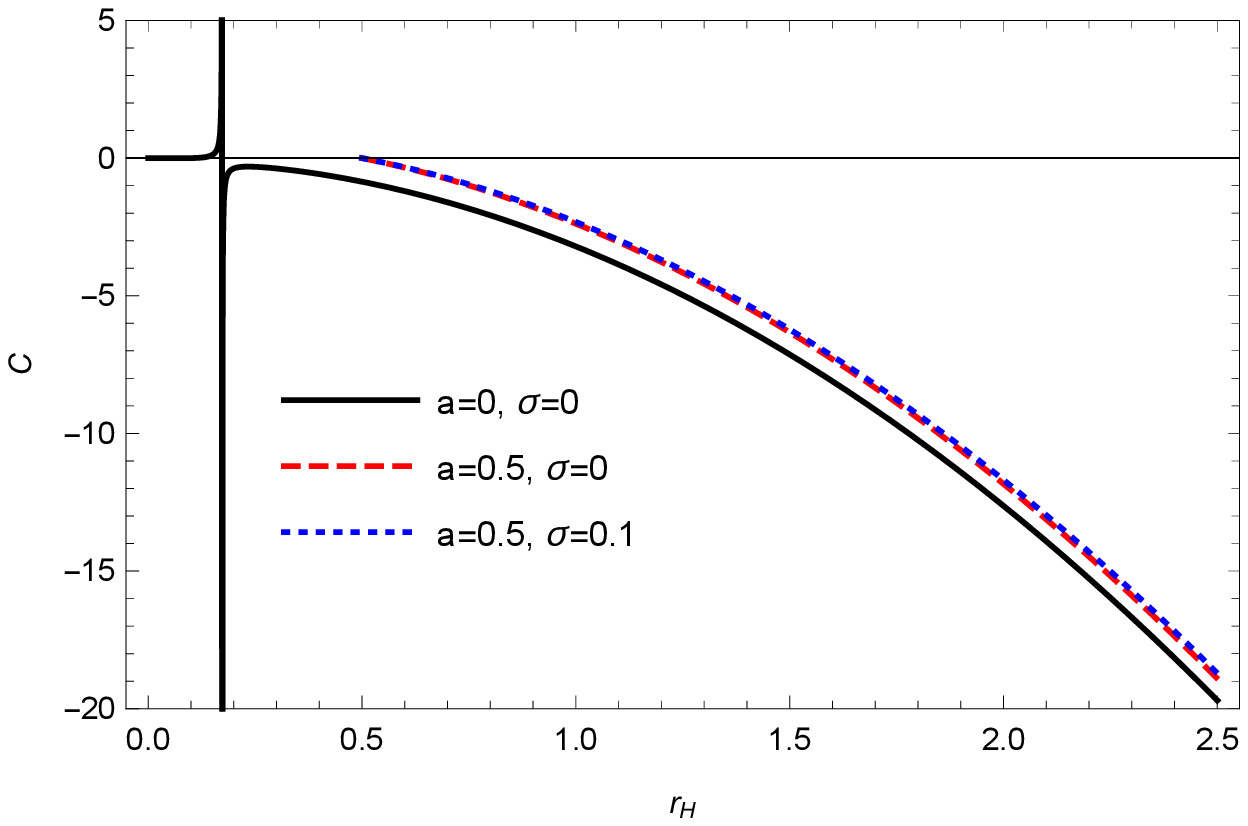}
       \subcaption{$ \omega_q=-0.35$, and $Q=0.1$.}
     \label{fig:Ca}
   \end{minipage}%
\begin{minipage}[t]{0.50\textwidth}
        \centering
       \includegraphics[width=\textwidth]{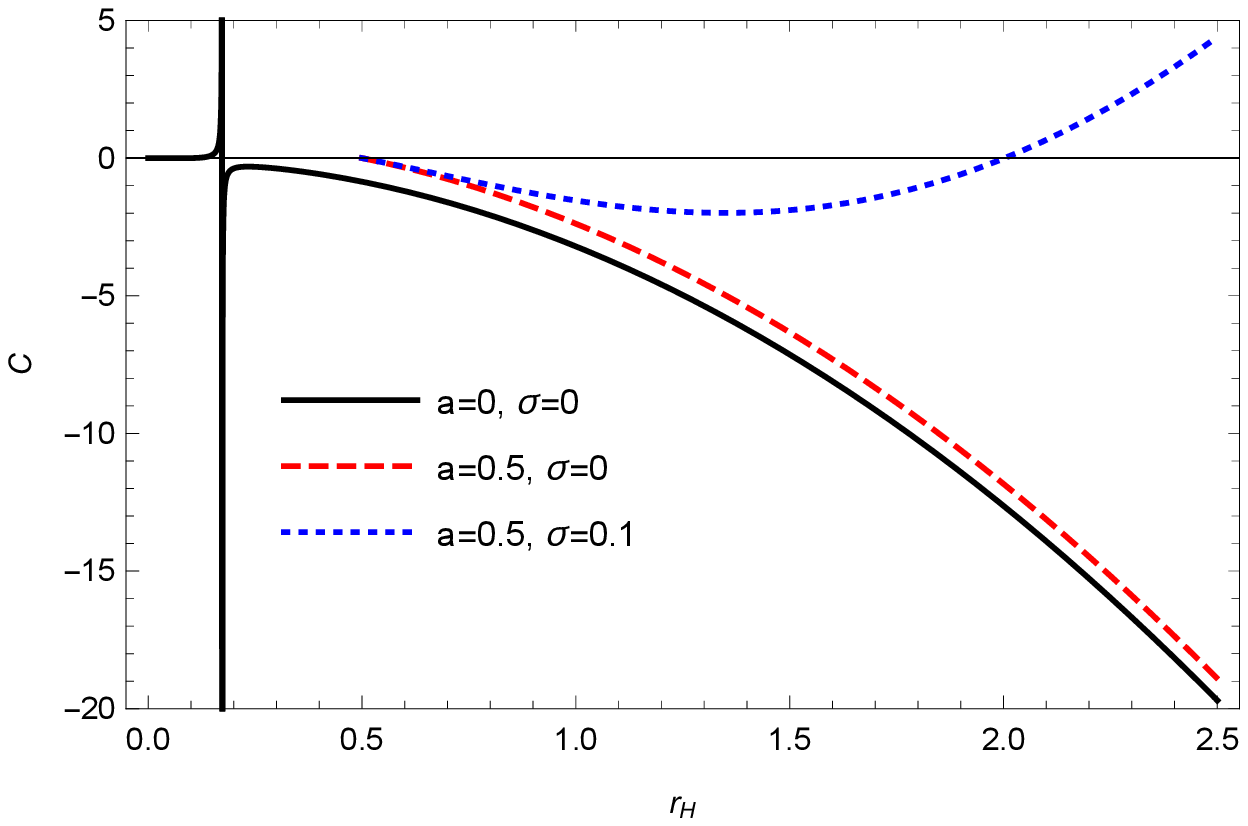}\\
        \subcaption{ $ \omega_q=-0.95$, and $Q=0.1$.}\label{fig:Cb}
    \end{minipage}\hfill
\begin{minipage}[t]{0.5\textwidth}
        \centering
        \includegraphics[width=\textwidth]{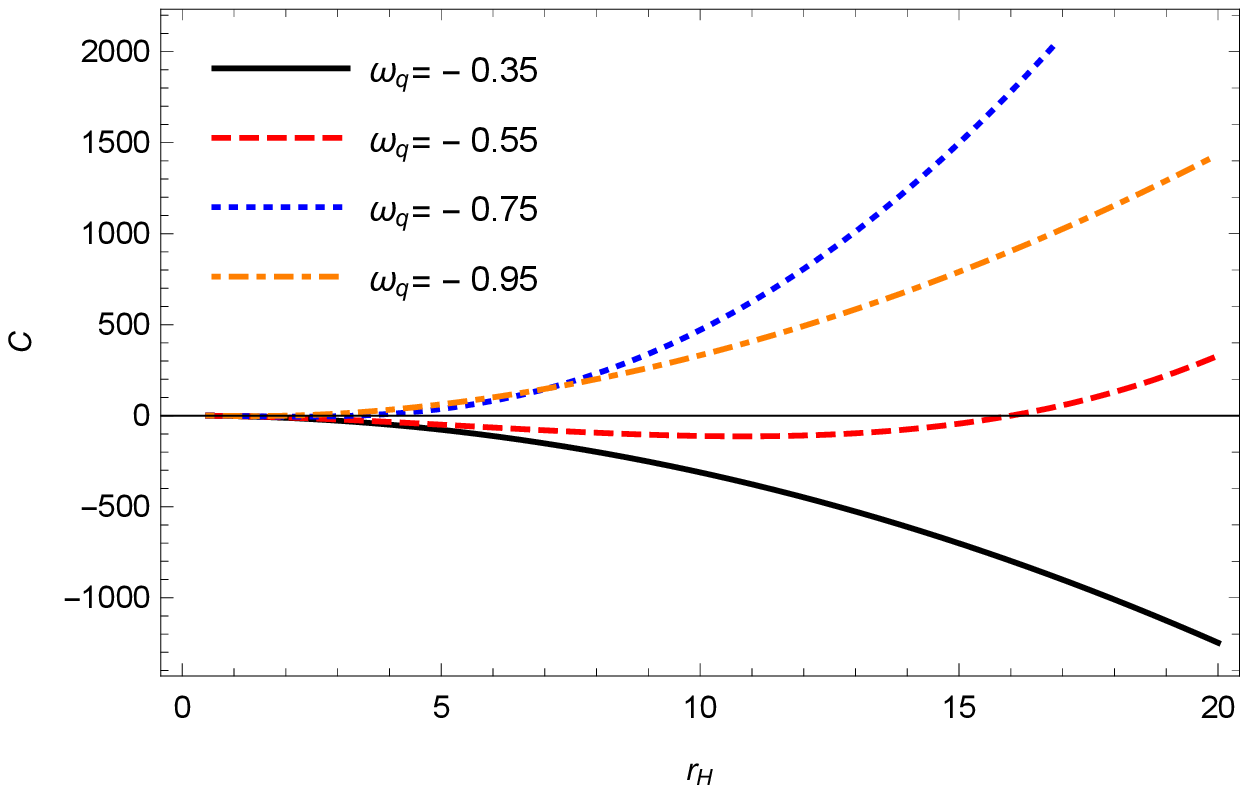}
       \subcaption{ $ a=0.5$, $Q=0.1$   and $\sigma=0.1$.}\label{fig:Cc}
   \end{minipage}%
\begin{minipage}[t]{0.50\textwidth}
        \centering
       \includegraphics[width=\textwidth]{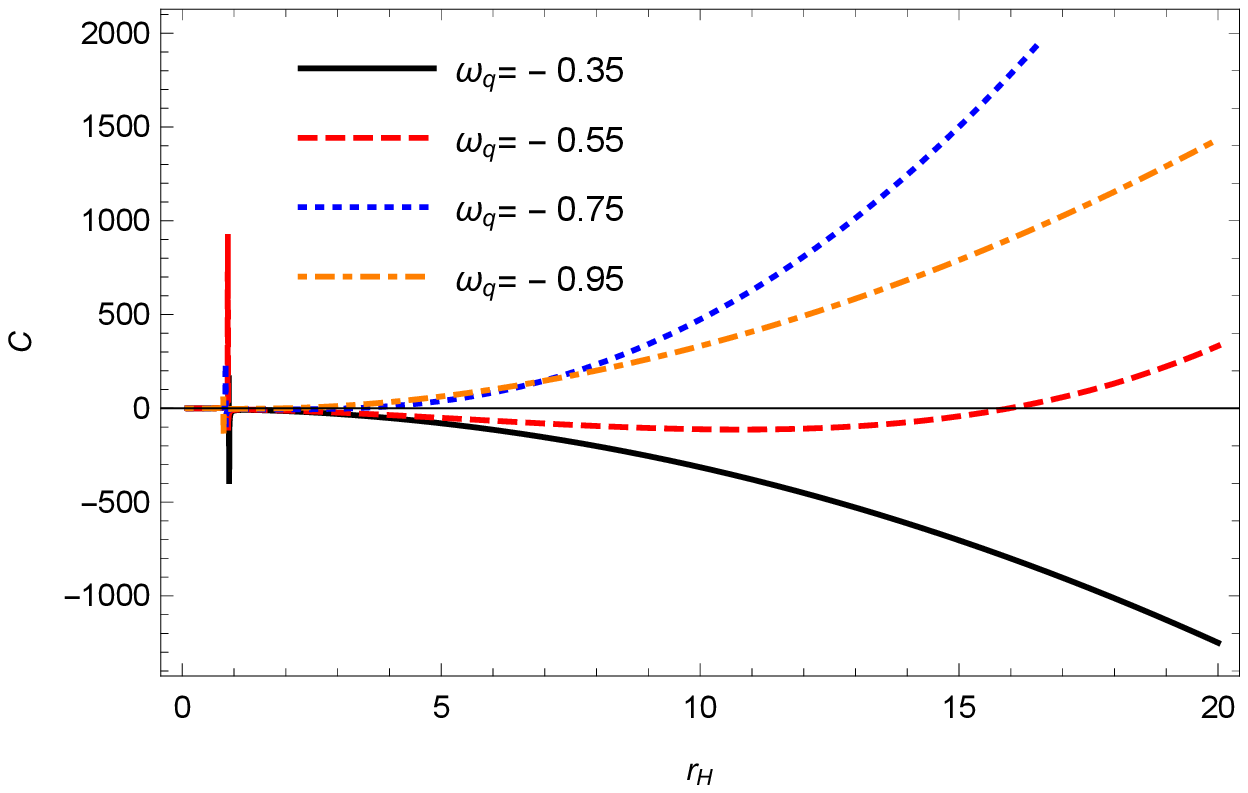}\\
        \subcaption{ $ a=0.1$, $Q=0.5$   and $\sigma=0.1$}\label{fig:Cd}
    \end{minipage}\hfill    
\caption{Quantum-corrected R-N black hole's heat capacity function versus event horizon radius in the presence of quintessence matter.}
\label{Cfigsnew}
\end{figure}

Next, we study the quantum-corrected heat capacity function via the well-known  thermodynamic relation:
\begin{equation}
C=\left( \frac{dM}{dT}\right){=\frac{dM}{d r_{H}}\left[\frac{dT_{H}}{d r_{H}}\right]^{-1}} .
\end{equation}
We obtain
{
\begin{equation}
C= - \pi r_{H}^{2} \Bigg(\frac{1}{\sqrt{1-\frac{a^{2}}{r_{H}^{2}}}}-\frac{Q^{2}}{r_{H}^{2}%
}+\frac{3\sigma \omega _{q}}{r_{H}^{3\omega _{q}+1}}\Bigg) \Bigg[
\frac{1}{\big( 1-\frac{a^{2}}{r_{H}^{2}}\big) ^{3/2}}-\frac{%
3Q^{2}}{r_{H}^{2}}+\frac{3\sigma \omega _{q}\left( 3\omega _{q}+2\right) }{%
r_{H}^{3\omega _{q}+1}}\Bigg]^{-1}.  \label{het}
\end{equation}%
}
To demonstrate the effects of the quintessence field and the quantum corrections we plot the quantum-corrected heat capacity function versus horizon in Fig. \ref{Cfigsnew}. 

Alike the previous plots, with Fig. \ref{fig:Ca} and Fig. \ref{fig:Cb} we present the influence of these modifications on the ordinary case. Then, in Fig. \ref{fig:Cc} and Fig. \ref{fig:Cd} we show $Q < a$ and $Q > a$ cases, respectively.  The black hole becomes unstable in larger intervals with smaller quintessence state parameter values.
{ Particularly, in the case of when $Q>a$,  the black hole with the horizon
radius $r_{H}<r_{\max }$  has the positive heat capacity, and thus it is
thermodynamically stable. Whereas, the black hole with the event radius $%
r_{rem}>r_{H}>r_{\max }$ black hole has negative heat capacity, and thereby it is
thermodynamically unstable. Moreover, the heat capacity is divergent at point $r_{H}=r_{\max}$, and changes its sign. This fact indicates that the black hole undergoes a second-order phase transition. In addition, at $r_{H}=r_{rem}$, the heat capacity is  equal to zero. This means that the black hole does not radiate, and thus, does not exchange heat with its surrounding. In the other case, where $Q<a$, we see that the black hole with the horizon radius $r_{H}<r_{rem}$, has negative heat capacity, thus the black hole is thermodynamically unstable, while for $r_{H}>r_{rem}$ the
black hole has the positive heat capacity, and so it is thermodynamically
stable. We observe also that for $r_{H}=r_{rem}$ the heat capacity tends to
zero.
}


\newpage

\begin{figure}[tbh]
    \begin{minipage}[t]{0.5\textwidth}
        \centering
        \includegraphics[width=\textwidth]{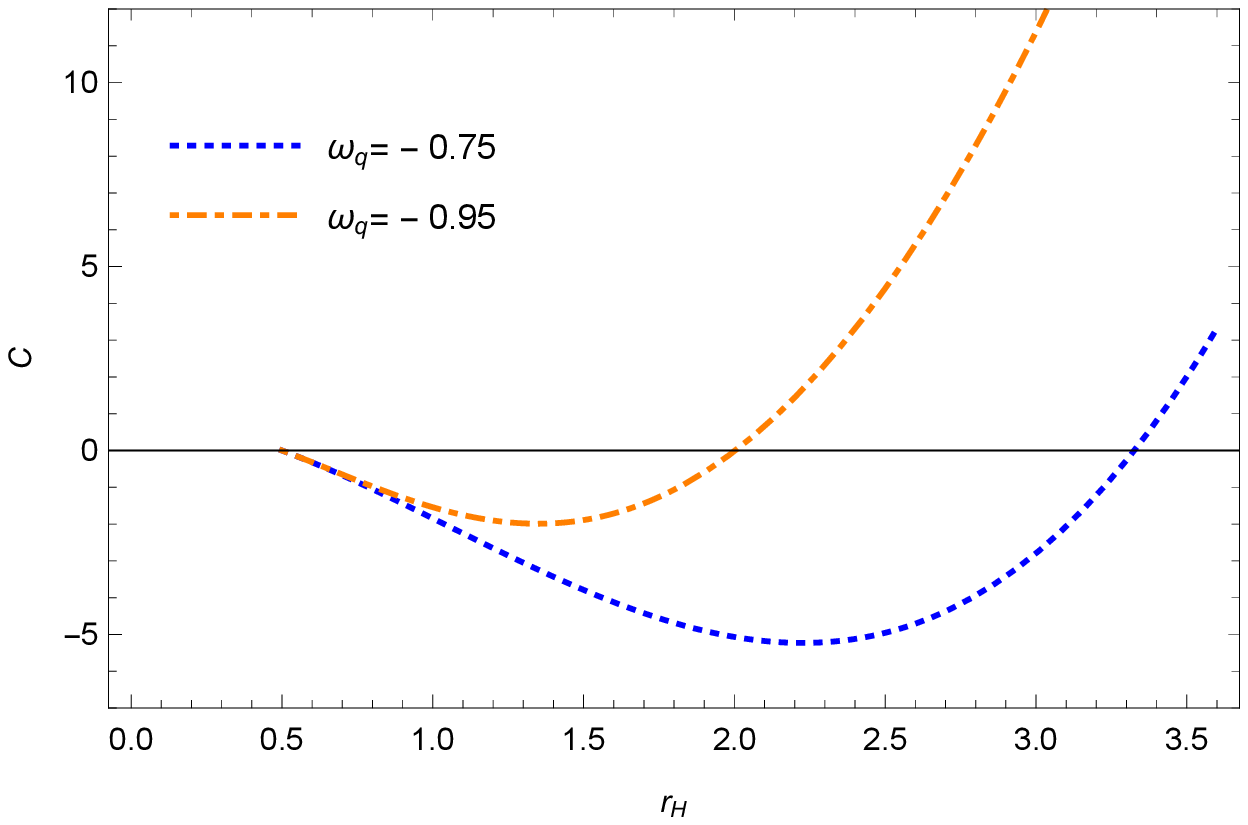}
       \subcaption {$a = 0.5 $, $Q = 0.1 $ and $\sigma = 0.1 $.}\label{fig:Ce}
   \end{minipage}%
\begin{minipage}[t]{0.50\textwidth}
        \centering
        \includegraphics[width=\textwidth]{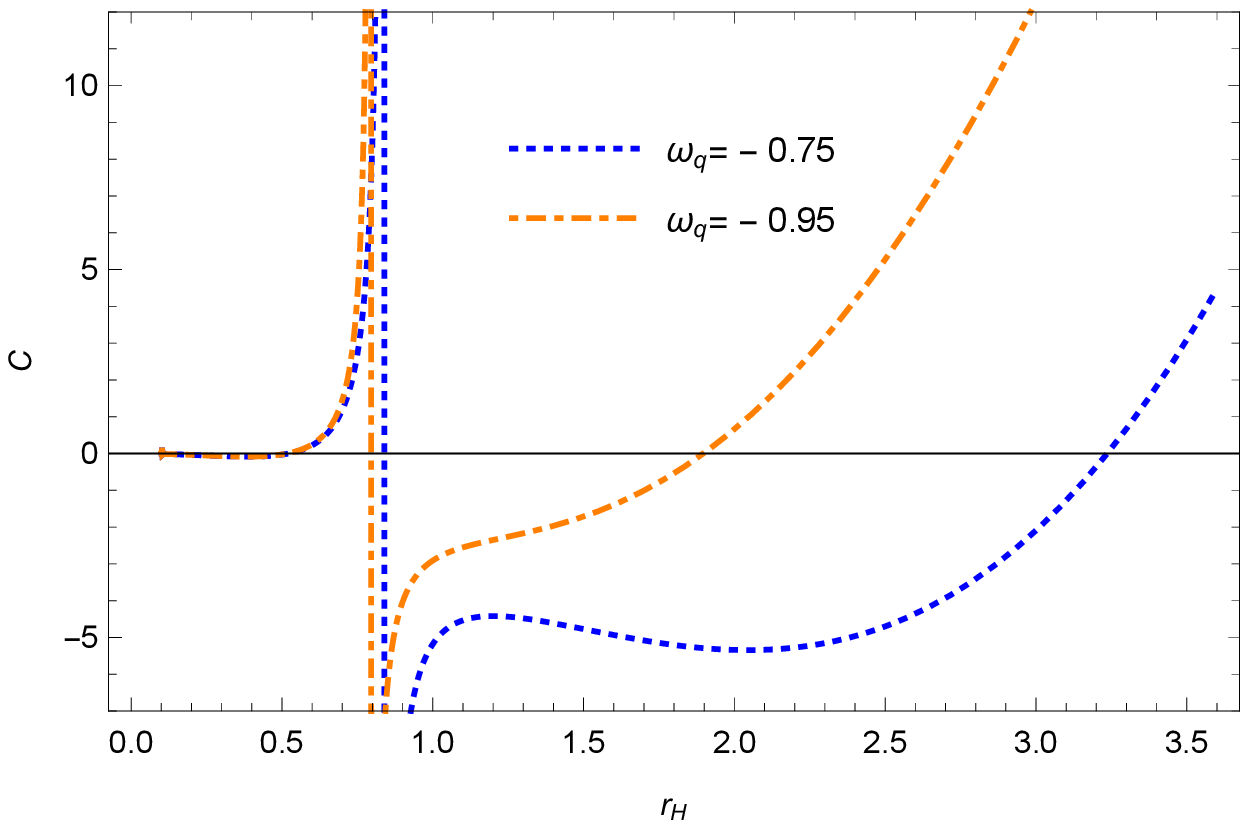}\\
        \subcaption {$a = 0.1 $, $Q = 0.5 $ and $\sigma = 0.1 $.}\label{fig:Cf}
    \end{minipage}\hfill
\caption{Quantum-corrected R-N black hole's heat capacity function versus event horizon radius in the presence of quintessence matter.}
\label{Cfigsnewek}
\end{figure}

For a more precise interpretation of the tables and discontinuity, we depict Fig. \ref{Cfigsnewek}.  In addition to the second second-order phase transitions, in Fig. \ref{fig:Ce} and Fig. \ref{fig:Cf} we observe first-order phase transitions at the points $r_H=r_{asy}$ where the specific heat is equal to zero.

\newpage
It is worth mentioning that the heat capacity function is an important tool to determine the black hole remnant mass. To state the existence of a remnant concretely,  one can take into account the zero values of the heat capacity. In Fig. \ref{fig:Ce}, for $\omega_q=-0.95$ and $\omega_q=-0.75$,  we see that R-N black holes are unstable in the physical interval. After $r_H = 2.003$ and $r_H = 3.326$, which are the remnant horizon radii, remnants occur with the given values in Table \ref{Tabfig2c}.  On the other hand, Fig. \ref{fig:Cf} shows that the black hole is stable between $r_{phys}$ and $r_{cr}$. Following the critical radius, the black hole becomes unstable. Therefore we conclude that the discontinuity corresponds to the critical radius where the black hole phase changes. Also, we notice that remnants occurs at $r_H = 1.897$ and $r_H = 3.236$ for $\omega_q=-0.95$ and $\omega_q=-0.75$ cases.  

{ We note that our results show some similarities with studies in the literature. For example, in \cite{Sudhaker}, Upadhyay et al. investigated the quantum fluctuation effects of the thermodynamic quantities of the B-T-Z black hole around thermal equilibrium for two special cases: charged non-rotating and uncharged rotating. They showed that quantum corrections only have a significant effect on black holes with relatively small horizon radii. In particular, with a negative deformation constant the black holes are always stable, however with a positive deformation constant case black holes are first unstable up to critical radii, and then stable after that value.  }

\begin{figure}[tbh]
\begin{minipage}[t]{0.5\textwidth}
        \centering
        \includegraphics[width=\textwidth]{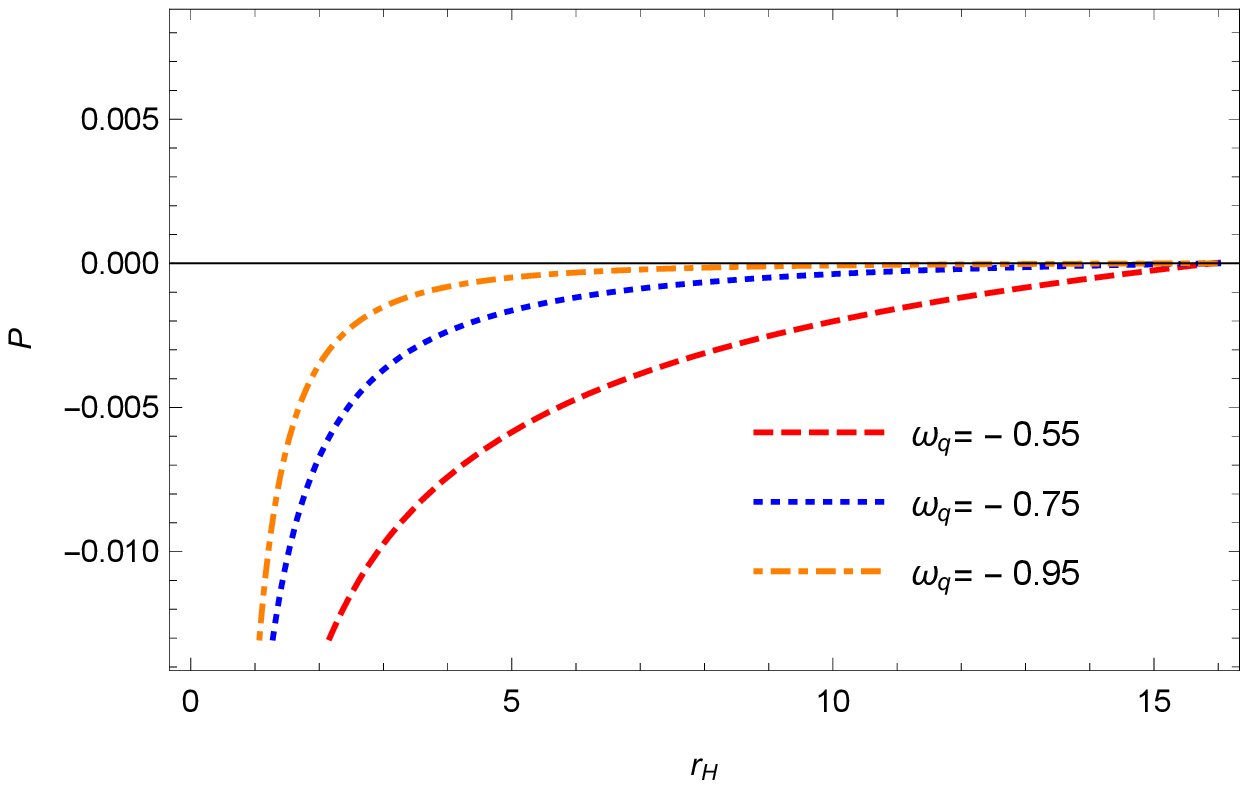}
       \subcaption{$T_H=0.05$, $a = 0.5 $, $Q = 0.1 $ and $\sigma = 0.1 $.}\label{fig:4a}
   \end{minipage}%
\begin{minipage}[t]{0.50\textwidth}
        \centering
        \includegraphics[width=\textwidth]{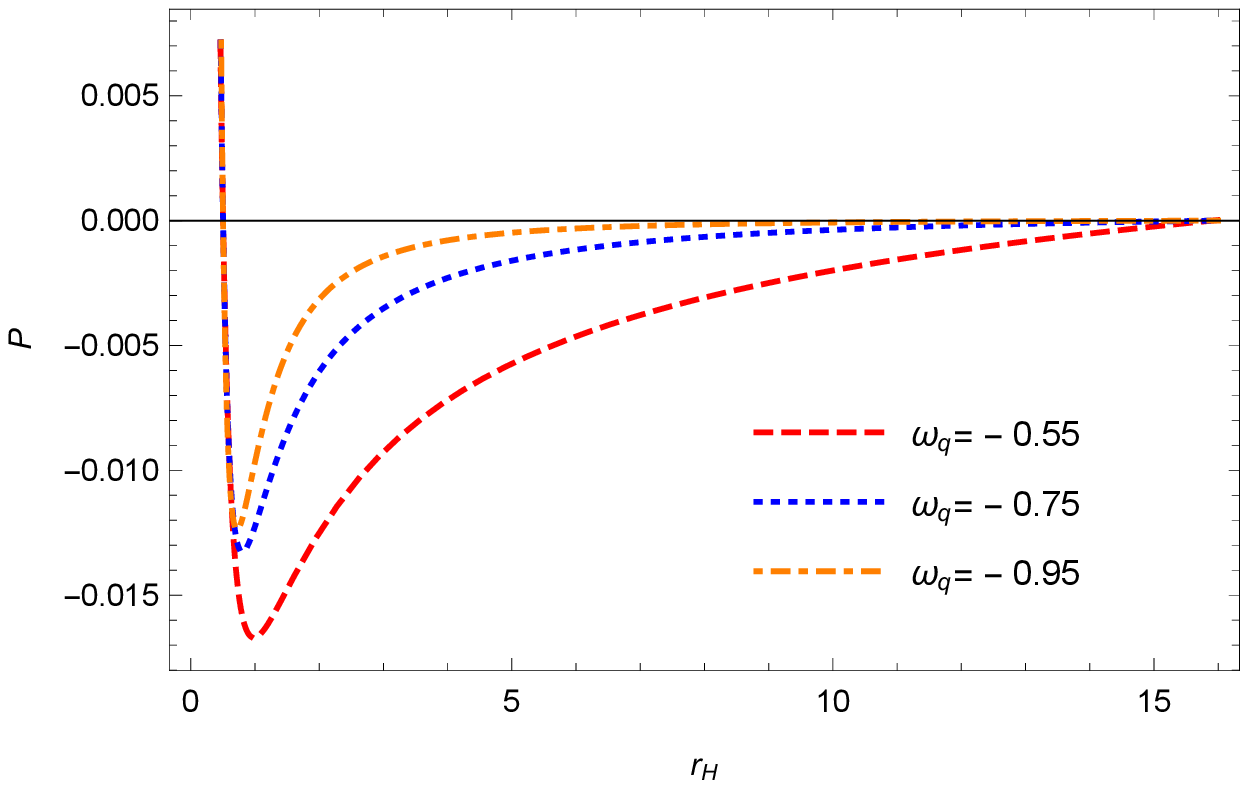}\\
        \subcaption{$T_H=0.05$, $a = 0.1 $, $Q = 0.5 $ and $\sigma = 0.1 $.}\label{fig:4b}
    \end{minipage}\hfill
\caption{Isotherms of quantum-corrected R-N black hole's equation of state function versus event horizon radius in the presence of quintessence matter.} \label{eqsrad}
\end{figure}

Now, let us focus on the study of the critical point of the quantum-corrected R-N black hole surrounded by quintessence. Employing the relation between pressure $P$ and $\sigma$ parameter %
\begin{equation}
P=-\frac{\sigma }{8\pi },  \label{R}
\end{equation}
we arrive at the equation of state $P=\left( T,r_{H}\right) $ in the form of
\begin{equation}
P=\frac{r_{H}^{3\omega_q+1}}{24 \pi \omega _{q}}\left( \frac{a^{2}}{r_{H}^{2}\sqrt{1-\frac{%
a^{2}}{r_{H}^{2}}}}+\sqrt{1-\frac{a^{2}}{r_{H}^{2}}}-\frac{Q^{2}}{r_{H}^{2}}%
-2\pi r_{H}T_{H}\right) .
\end{equation}
We depict the isotherms of the equation of state function versus horizon radius in Fig. \ref{eqsrad} for $Q < a$ and $Q > a$ cases, respectively. 

{
At the end of this section, we focus on the study of criticality. We note that critical points can be derived from solutions to the following conditions:
\begin{eqnarray}
\left. \frac{\partial P}{\partial r_{H}}\right\vert _{T_{c},r_{c}}&=&0, \\ 
\left. \frac{\partial ^{2}P}{\partial r_{H}^{2}}\right\vert _{T_{c},r_{c}}&=&0.
\end{eqnarray}%
First, we Taylor expand the equation of state function for small values of the quantum correction parameter $a<<1$. We arrive at the following form:%
\begin{equation}
P=\frac{r_{H}^{3\omega _{q}+1}}{24\pi \omega _{q}}\left( 1-\frac{Q^{2}-\frac{%
a^{2}}{2}}{r_{H}^{2}}-2\pi r_{H}T_{H}\right) .
\end{equation}
Then, using the first constraint, we obtain the following expressions for the critical values of $T_{c}$ and $P_{c}$ for an arbitrary state parameter $\omega _{q}$.
\begin{eqnarray}
T_{c}&=&\frac{1}{2\pi r_{c}\left( 3\omega _{q}+2\right) }\left[ 1+3\omega _{q}-%
\frac{\left( Q^{2}-\frac{a^{2}}{2}\right) \left( 3\omega _{q}-1\right) }{%
r_{c}^{2}}\right] , \\
P_{c}&=&\frac{r_{c}^{3\omega _{q}+1}}{24\pi \omega _{q}\left( 3\omega
_{q}+2\right) }\left( 1-\frac{3\left( Q^{2}-\frac{a^{2}}{2}\right) }{%
r_{c}^{2}}\right) ,
\end{eqnarray}
where the critical radius $r_{c}$ is the solution of the following equation:%
\begin{equation}
\left( 9\omega _{q}^{2}-9\omega _{q}+2\right) \left( Q^{2}-\frac{a^{2}}{2}%
\right) -3\omega _{q}r_{c}^{2}\left( 3\omega _{q}+1\right) +2\pi \left(
9\omega _{q}^{2}+9\omega _{q}+2\right) T_{c}r_{c}^{3}=0.
\end{equation}%
After simple algebra, we find the following critical values: 
\begin{eqnarray}
r_{c}&=&\sqrt{\left( Q^{2}-\frac{a^{2}}{2}\right) \left( \frac{9\omega _{q}-3}{%
3\omega _{q}+1}\right) }, \\
P_{c}&=&-\frac{\left( Q^{2}-\frac{a^{2}}{2}\right) ^{\frac{3\omega _{q}+1}{2}}%
}{12\pi \omega _{q}\left( 3\omega _{q}+2\right) \left( 3\omega _{q}-1\right) 
}\left( \frac{9\omega _{q}-3}{3\omega _{q}+1}\right) ^{\frac{3\omega _{q}+1}{%
2}}, \\
T_{c}&=&\frac{3\omega _{q}+1}{3\pi \left( 3\omega _{q}+2\right) }\sqrt{\frac{%
3\omega _{q}+1}{\left( Q^{2}-\frac{a^{2}}{2}\right) \left( 9\omega
_{q}-3\right) }}.
\end{eqnarray}%
We notice that to ensure a positive critical pressure, temperature, and radius the following conditions must be fulfilled:%
\begin{equation}
Q>\frac{a}{\sqrt{2}},\text{ \ \ and \ \ \ }-1<\omega _{q}<-2/3.
\end{equation}%
Then, by defining the quantity%
\begin{equation}
v_{c}=-\frac{3\omega _{q}\left( 9\omega _{q}^{2}-1\right) }{2r^{3\omega
_{q}+2}},
\end{equation}%
with straightforward algebra, we find a dimensionless golden relation for the
quantities $P_{c}$, $v_{c}$ and $T_{c}$.%
\begin{equation}
\frac{P_{c}v_{c}}{T_{c}}=\frac{3}{8},
\end{equation}%
which is similar to relations found in the context of the Van der Waals
equation. This relation depends on the kind of fluid that surrounds the
black hole, and the scale of the quantum correction.
}

\section{Shadow of the black hole}

Our next step is to examine the shadow of the quantum-corrected R-N black hole surrounded by quintessence matter. Since the metric possesses spherical symmetry, we can express the Lagrangian according to the given metric coefficient in Eq. \eqref{met} as follows:
\begin{equation}
\mathcal{L}\left( q,\dot{q}\right) =\frac{1}{2}\left[ -f\left( r\right) \dot{%
t}^{2}+\frac{1}{f\left( r\right) }\dot{r}^{2}+r^{2}\dot{\theta}%
^{2}+r^{2}\sin ^{2}\theta \dot{\varphi}^{2}\right] .
\end{equation}%
Here, the dot over the variables signifies the differentiation with respect to proper time, $\lambda $. Then, we find the canonically conjugate momentum in the form of
\begin{eqnarray}
P_{t}&=&\left( \sqrt{1-\frac{a^{2}}{r^{2}}}-\frac{2M}{r}+\frac{Q^{2}}{r^{2}}-%
\frac{\sigma }{r^{3\omega _{q}+1}}\right) \dot{t}=E,  \label{a} \\
P_{\varphi }&=&r^{2}\sin ^{2}\theta \dot{\varphi}=L,  \label{b} \\
P_{\theta }&=&r^{2}\dot{\theta},  \label{c} \\
P_{r}&=&\frac{1}{f\left( r\right) }\dot{r}.  \label{d}
\end{eqnarray}%
Here we have the two important constants of motion: $E$ and $L$, the energy and the angular momentum of the test particle. To investigate photon orbits around the black hole,  we adopt the Hamilton-Jacobi method and employ the formulation of geodesic equations by the Carter approach for quantum-corrected R-N black hole spacetime \cite{Carter}. To this end, we consider a separable solution of the Jacobi action, {$\mathcal{S}$},
\begin{equation}
{\mathcal{S}}=-Et+L\varphi +{\mathcal{S}_{r}}\left( r\right) +{\mathcal{S}_{\theta }}\left( \theta \right) , \label{sep}
\end{equation}
where {$\mathcal{S}_{r}\left( r\right) $ and $\mathcal{S}_{\theta }\left( \theta \right) $} are the functions of $r$ and $\theta$. Then, by substituting Eqs. (\ref{L}) and (\ref{sep}) into the Hamilton-Jacobi equation of the form
\begin{equation}
\frac{\partial {\mathcal{S}}}{\partial \lambda }+\frac{1}{2}g^{\mu \nu }\frac{\partial {\mathcal{S}}%
}{\partial x^{\mu }}\frac{\partial {\mathcal{S}}}{\partial x^{\nu }}=0,  \label{hj}
\end{equation}%
we get
{\small\begin{eqnarray}
{r^4}\left( \sqrt{1-\frac{a^{2}}{r^{2}}}-\frac{2M}{r}+\frac{Q^{2}}{r^{2}}-%
\frac{\sigma }{r^{3\omega _{q}+1}}\right) ^{2}\left( \frac{\partial {\mathcal{S}_{r}}}{%
\partial r}\right) ^{2}&=&{r^4E^{2}-r^2}\left( \sqrt{1-\frac{a^{2}}{r^{2}}}-%
\frac{2M}{r}+\frac{Q^{2}}{r^{2}}-\frac{\sigma }{r^{3\omega _{q}+1}}\right) %
\left[ \mathcal{K}+L^{2}\right] , \,\,\,\,\,\,\,\,\,  \label{27} \\
\left( \frac{\partial {\mathcal{S}_{\theta }}}{\partial \theta }\right) ^{2}&=&\mathcal{K}%
-L^{2}\cot ^{2}\theta ,  \label{25}
\end{eqnarray}}\normalsize
where $\mathcal{K}$ is called the Carter constant \cite{Carter}. Then, we recast Eqs. \eqref{27} and \eqref{25} with the following  separated equations
\begin{equation}
r^{2}\frac{d\theta }{d\lambda }=\pm \sqrt{\Theta }=\pm \sqrt{\mathcal{K}%
-L^{2}\cot ^{2}\theta },
\end{equation}%
\begin{equation}
r^{2}\frac{dr}{d\lambda }=\pm \sqrt{\mathcal{R}}=\pm \sqrt{%
r^{4}E^{2}-r^{2}\left( \sqrt{1-\frac{a^{2}}{r^{2}}}-\frac{2M}{r}+\frac{Q^{2}%
}{r^{2}}-\frac{\sigma }{r^{3\omega _{q}+1}}\right) \left[ \mathcal{K}+L^{2}%
\right] }.
\end{equation}
We notice that the latter equation can be expressed in the following familiar form%
\begin{equation}
\left( \frac{dr}{d\lambda }\right) ^{2}+V_{eff}\left( r\right) =0, \label{cann}
\end{equation}%
with the effective potential 
\begin{equation}
V_{eff}\left( r\right) =\frac{\mathcal{K}+L^{2}}{r^{2}}\left( \sqrt{1-\frac{%
a^{2}}{r^{2}}}-\frac{2M}{r}+\frac{Q^{2}}{r^{2}}-\frac{\sigma }{r^{3\omega_{q}+1}}\right) -E^{2}.
\end{equation}
It is worth noting that Eq. \eqref{cann} describes the radial motion of the particle that depends on the impact parameters $\xi =\frac{L}{E}$, and $\eta =\frac{\mathcal{K}}{E^{2}}$
which has the same form of energy conservation law in one-dimensional classical mechanics. Here, $\lambda $ acts as the time variable.

According to the impact parameters, $\eta $ and $\xi $,  Chandrasekhar ranked the photon orbits into three groups: scattering, plunging, and unstable circular and spherical orbits \cite{Kumar}.  The unstable orbits set apart the plunging and scattering orbits with the radii given in \cite{Chandrasekhar} by
\begin{equation}
\left. V_{eff}\left( r\right) \right\vert _{r=r_{p}}=\left. \frac{dV_{eff}\left( r\right) }{dr}\right\vert _{r=r_{p}}=0,
\end{equation}
or 
\begin{equation}
\left. \mathcal{R}\right\vert _{r=r_{p}}=\left. \frac{d\mathcal{R}}{dr}%
\right\vert _{r=r_{p}}=0,
\end{equation}
where $r_{p}$ is the radius of the photon sphere. By solving the two equations simultaneously, we find the following equation for the radius of a photon sphere:
\begin{equation}
r_{p}f^{\prime }\left( r_{p}\right) -2f\left( r_{p}\right) =0,
\end{equation}
which does not have an analytical solution, and thus, numerical methods have to be used to obtain the roots. In so doing, we generate Table \ref{ShadowTable} to present  numerical results of the event horizon, the radius of the photon sphere, and the impact parameters of the photon sphere for different values of $Q,$ $a$, and $\omega _{q}$ for $M=0.1$ and $\sigma =0.05$.
\begin{table}[tbp]
\center%
\begin{tabular}{|l|l|l|l|l|l|}
\hline
$\omega _{q}$ & $Q$ & $a$ & $r_{H}$ & $r_{p}$ & $\eta +\xi ^{2}$ \\ \hline
$-0.4$ & $0.05$ & $0$ & $0.194113$ & $0.293278$ & $0.279263$ \\ 
& $0.05$ & $0.02$ & $0.195266$ & $0.294788$ & $0.281368$ \\ 
& $0.05$ & $0.03$ & $0.196696$ & $0.296661$ & $0.28399$ \\ 
& $0.05$ & $0.04$ & $0.198675$ & $0.299254$ & $0.287643$ \\ \hline $-0.6$   & $0.05$ & $0$ & $0.189292$ & $0.28564$ & $0.261415$ \\ 
& $0.05$ & $0.02$ & $0.190453$ & $0.287158$ & $0.263465$ \\ 
& $0.05$ & $0.03$ & $0.191893$ & $0.28904$ & $0.266018$ \\ 
& $0.05$ & $0.04$ & $0.193886$ & $0.291646$ & $0.269577$ \\ \hline
$-0.8$ & $0.05$ & $0$ & $0.187574$ & $0.283046$ & $0.253457$ \\ 
& $0.05$ & $0.02$ & $0.18873$ & $0.284573$ & $0.255476$ \\ 
& $0.05$ & $0.03$ & $0.190166$ & $0.286446$ & $0.257965$ \\ 
& $0.05$ & $0.04$ & $0.192153$ & $0.28904$ & $0.261435$ \\ \hline
\end{tabular}%
\caption{The values of event horizon, $r_{H}$, photon radius, $r_{o}$, and
impact parameters $\protect\eta +\protect\xi ^{2}$, for different values of $%
Q,$ $a$ and $\protect\omega _{q}$, with $M=0.1$ , $\protect\sigma =0.05$. } \label{ShadowTable}
\end{table}

\begin{figure}[tbh]
\begin{minipage}[t]{0.5\textwidth}
        \centering
        \includegraphics[width=\textwidth]{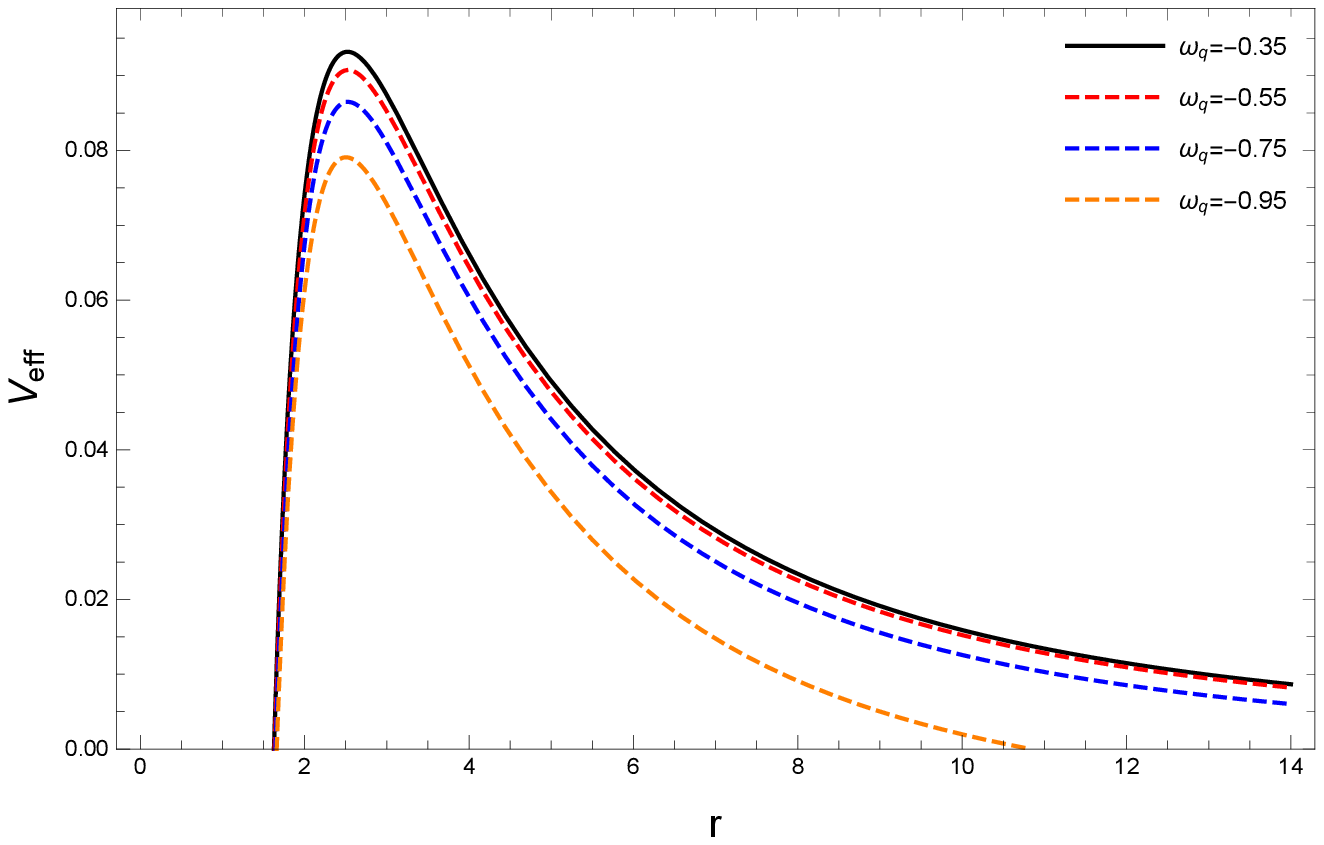}
       \subcaption{$a=0.1$, $Q=0.8$,  $L=1$  and $\sigma=0.01$.}  \label{fig:Pota}
   \end{minipage}%
\begin{minipage}[t]{0.50\textwidth}
        \centering
       \includegraphics[width=\textwidth]{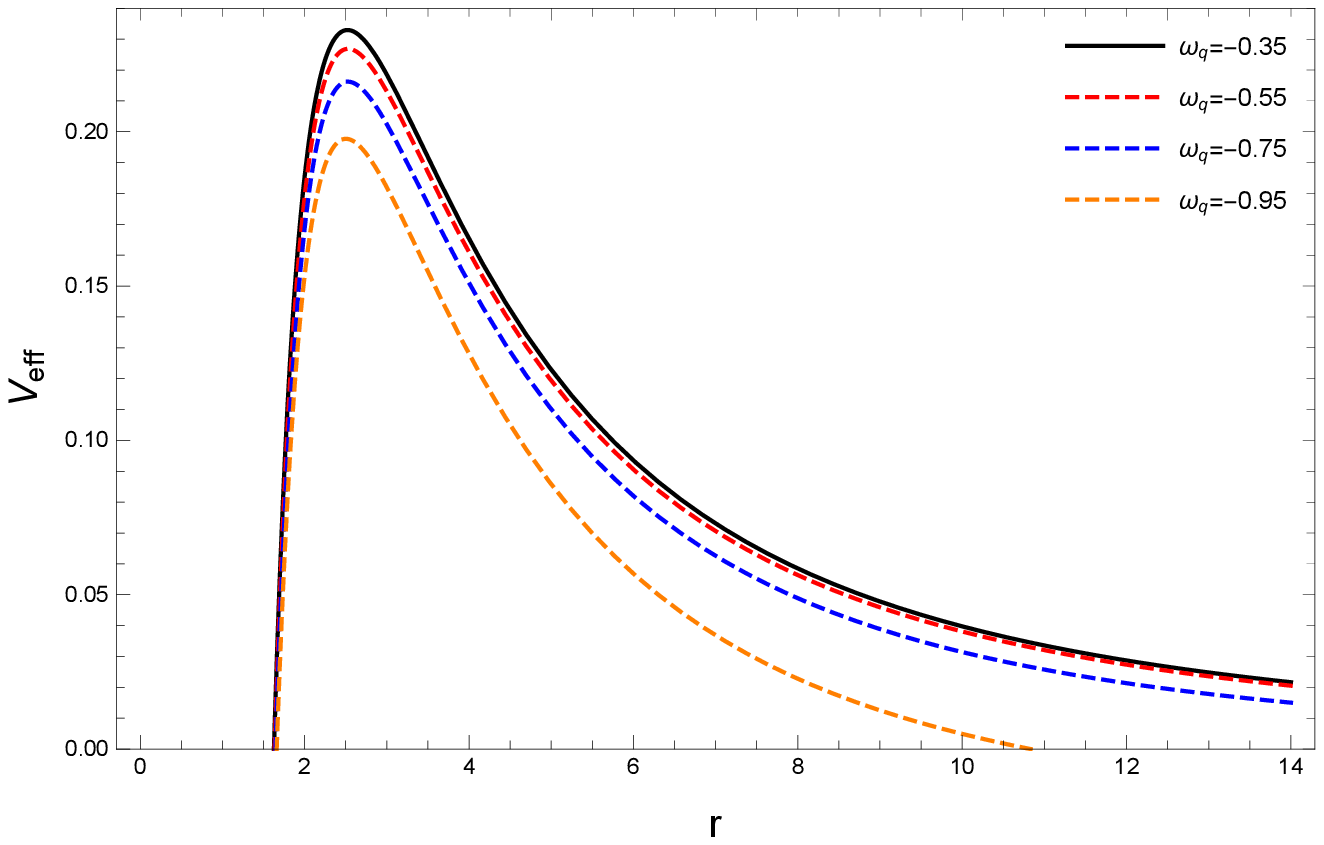}
        \subcaption{ $ a=0.1$, $Q=0.8$,  $L=2$  and $\sigma=0.01$.}\label{fig:Potb}
    \end{minipage}\hfill
\begin{minipage}[t]{0.5\textwidth}
        \centering
        \includegraphics[width=\textwidth]{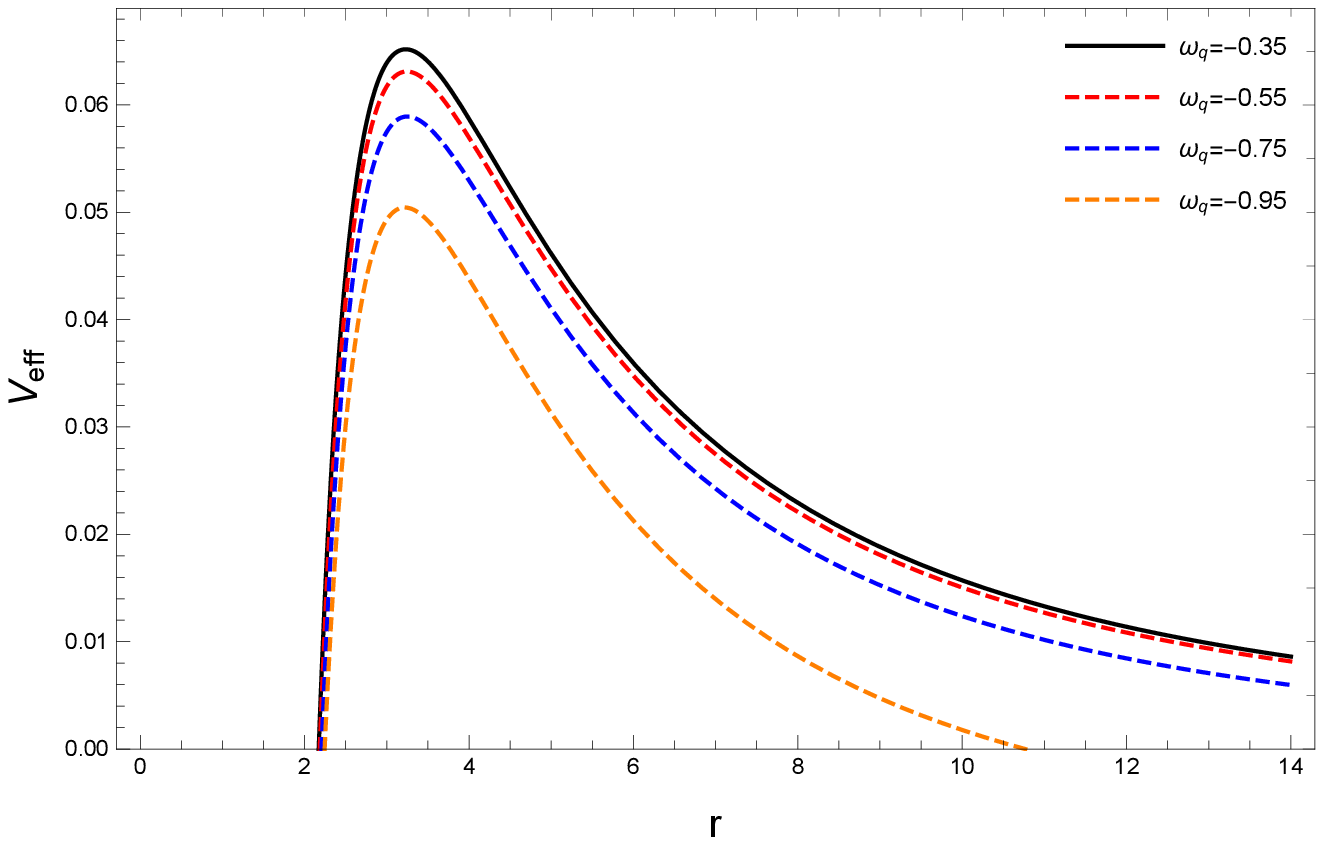}
       \subcaption{$a=0.8$, $Q=0.1$, $L=1$ and $\sigma=0.01$.}\label{fig:Potc}
   \end{minipage}%
\begin{minipage}[t]{0.50\textwidth}
        \centering
       \includegraphics[width=\textwidth]{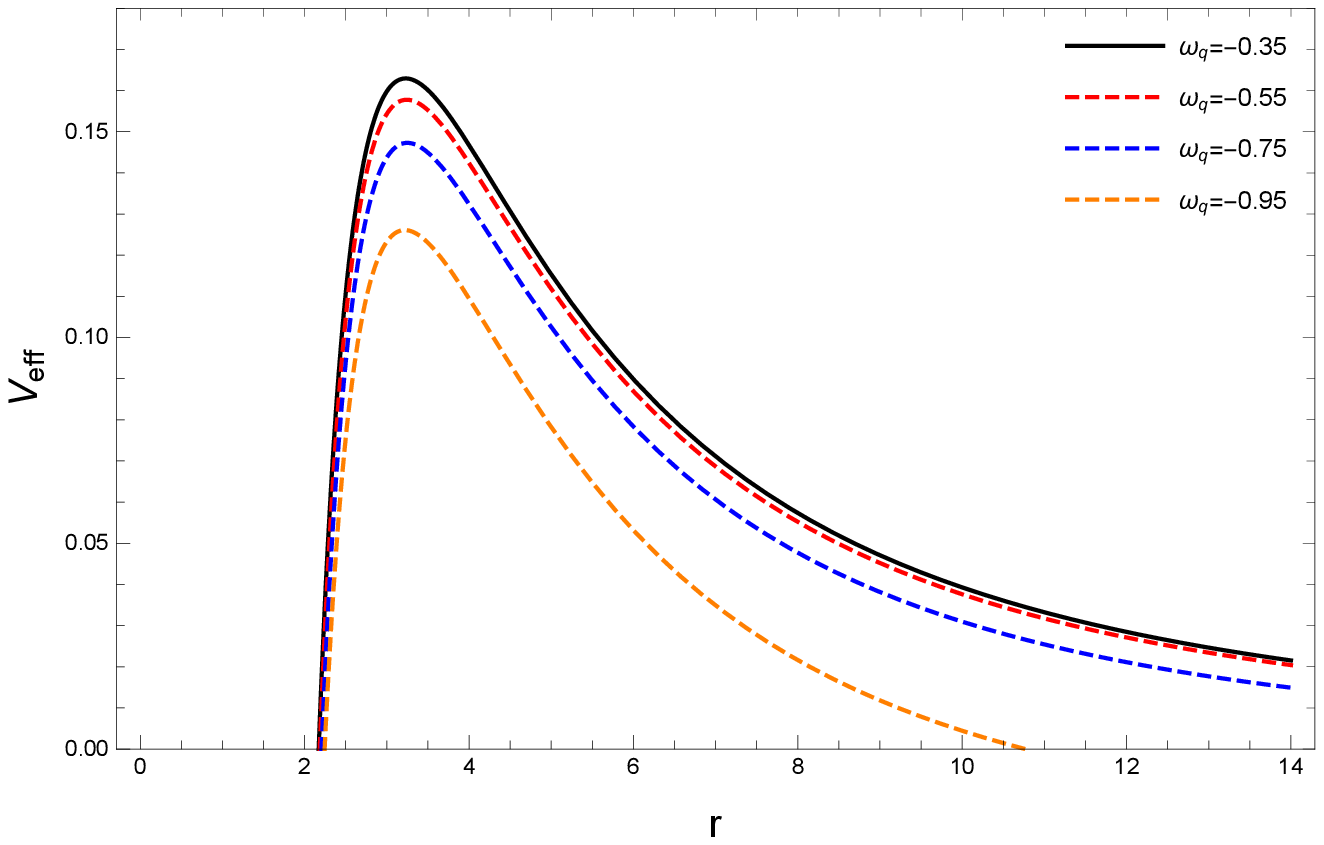}
        \subcaption{ $a=0.8$, $Q=0.1$, $L=2$ and $\sigma=0.01$.}\label{fig:Potd}
    \end{minipage}\hfill    
\caption{The effective potential as a function of the radial coordinate $r$ with different values of $Q$, $a$, $L$, and $\omega _{q}$ for 
 {$\mathcal{K}=M=1.$} }\label{Effectivepot}
\end{figure}

\newpage
For $Q>a$ and $Q<a$ cases, we depict the effective potential versus radial coordinate in Fig. \eqref{Effectivepot}. In both cases, we observe that angular momentum strength and the quintessence state parameter alter the strength of the effective potential.

\begin{figure}[tbh]
\begin{minipage}[t]{0.5\textwidth}
        \centering
        \includegraphics[width=\textwidth]{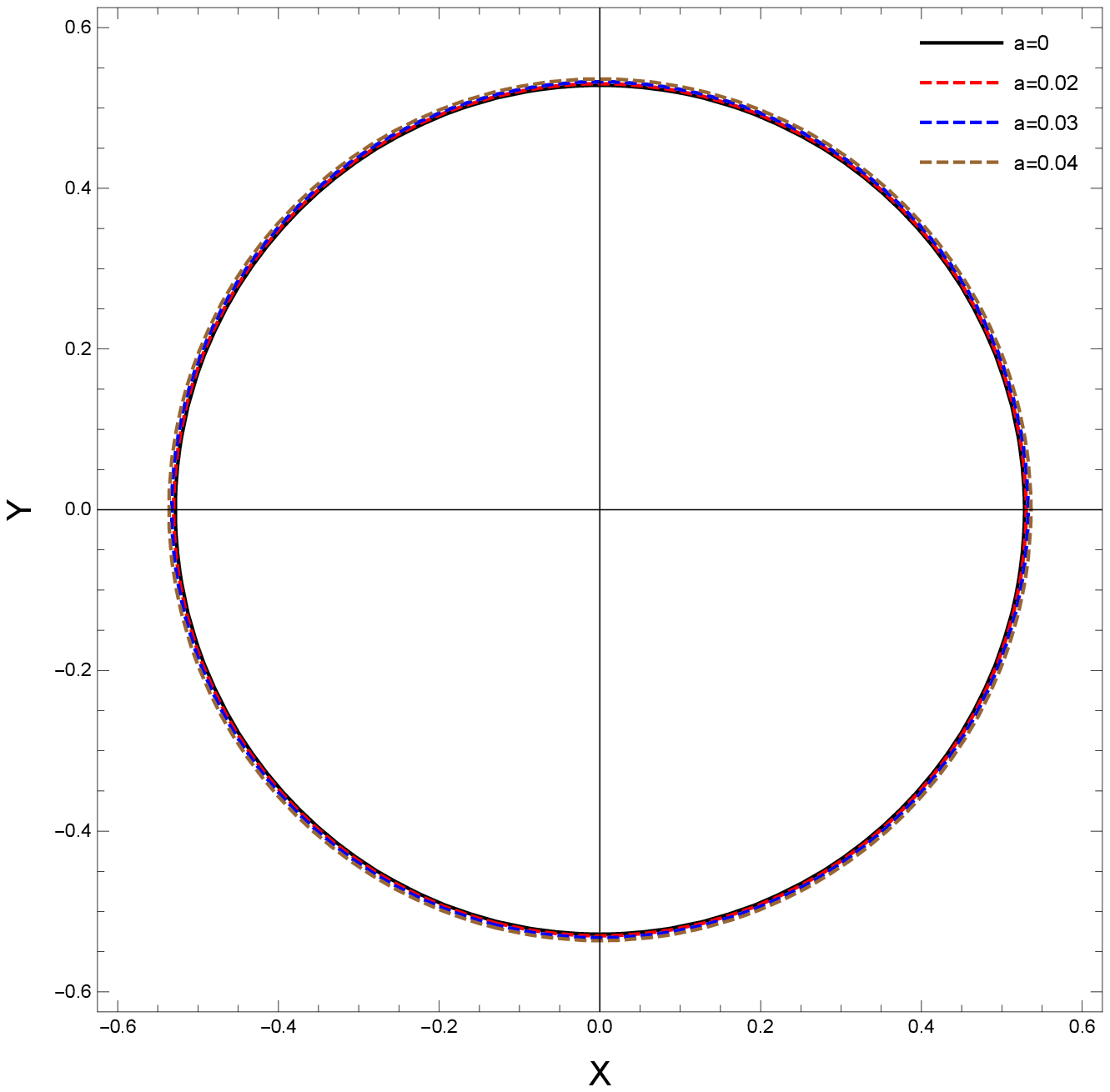}
       \subcaption{$Q=0.05$, and $\omega_{q}=-0.4$.}
     \label{fig:Shadowa}
   \end{minipage}%
\begin{minipage}[t]{0.5\textwidth}
        \centering
       \includegraphics[width=\textwidth]{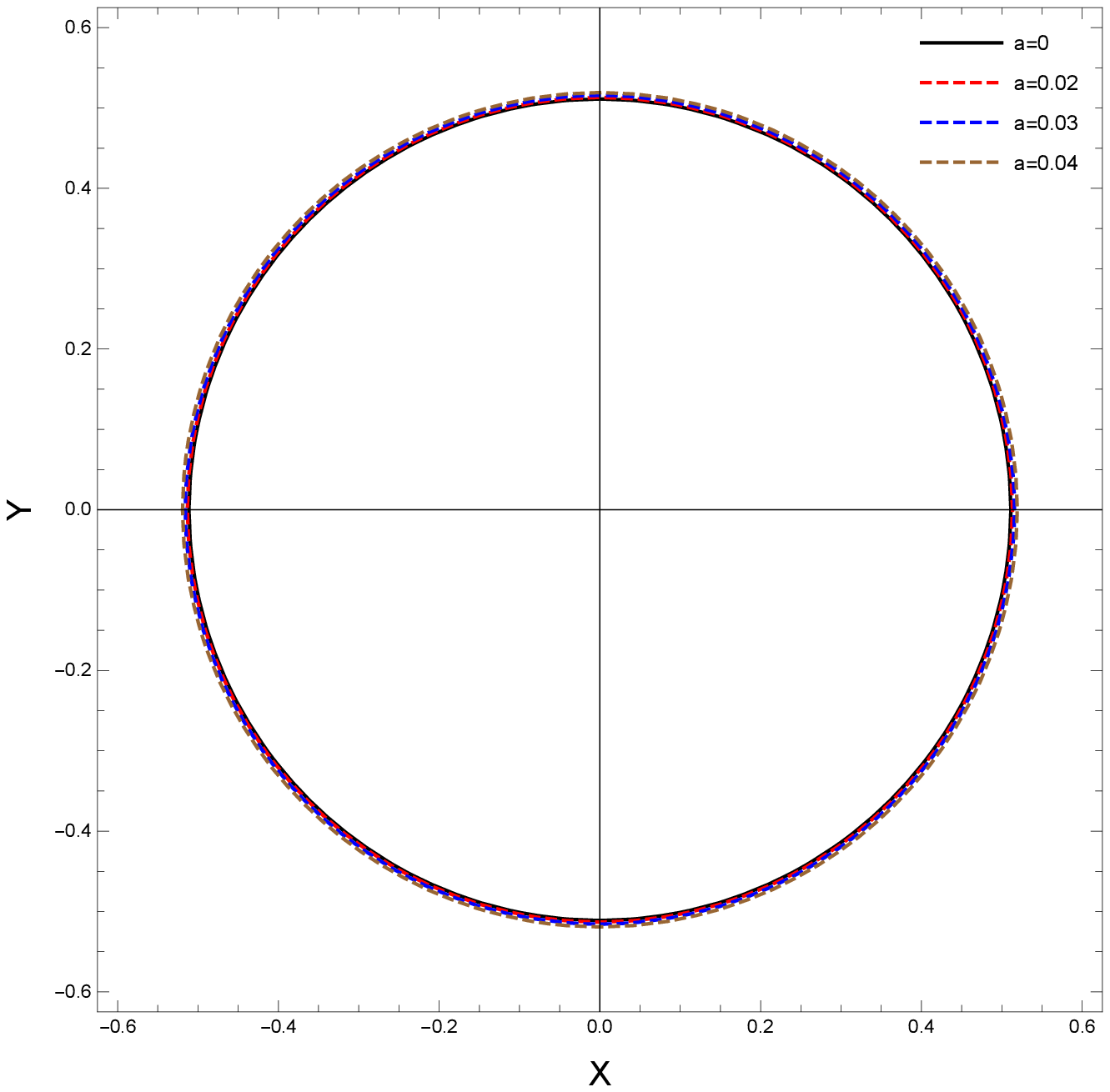}\\
        \subcaption{$Q=0.05$, and $\omega_{q}=-0.6$.}\label{fig:Shadowb}
    \end{minipage}\hfill
\begin{minipage}[t]{0.5\textwidth}
        \centering
        \includegraphics[width=\textwidth]{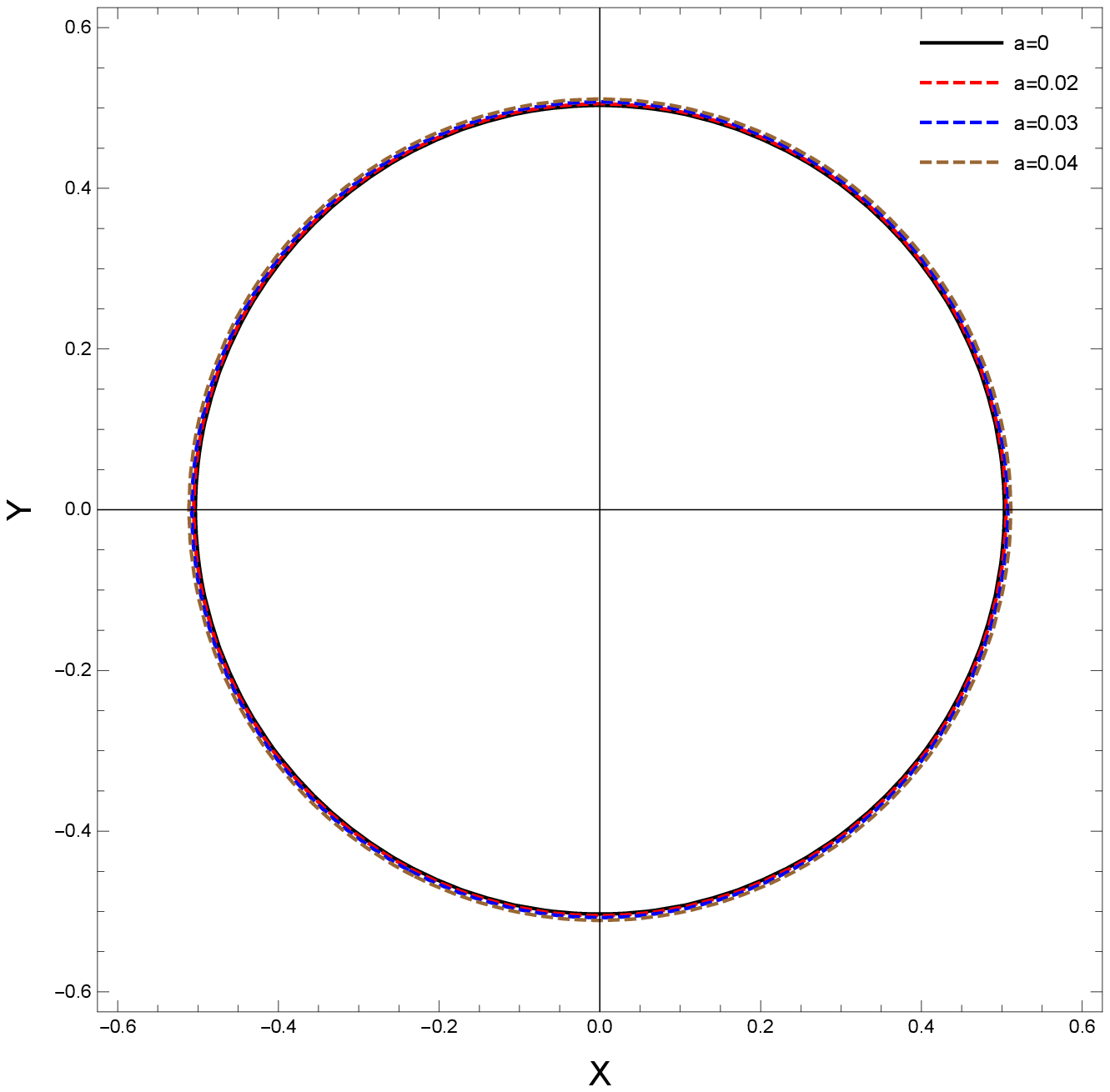}
       \subcaption{$a=0.8$, $Q=0.1$, $L=1$ and $\sigma=0.01$.}\label{fig:Shadowc}
   \end{minipage}%
\begin{minipage}[t]{0.5\textwidth}
        \centering
       \includegraphics[width=\textwidth]{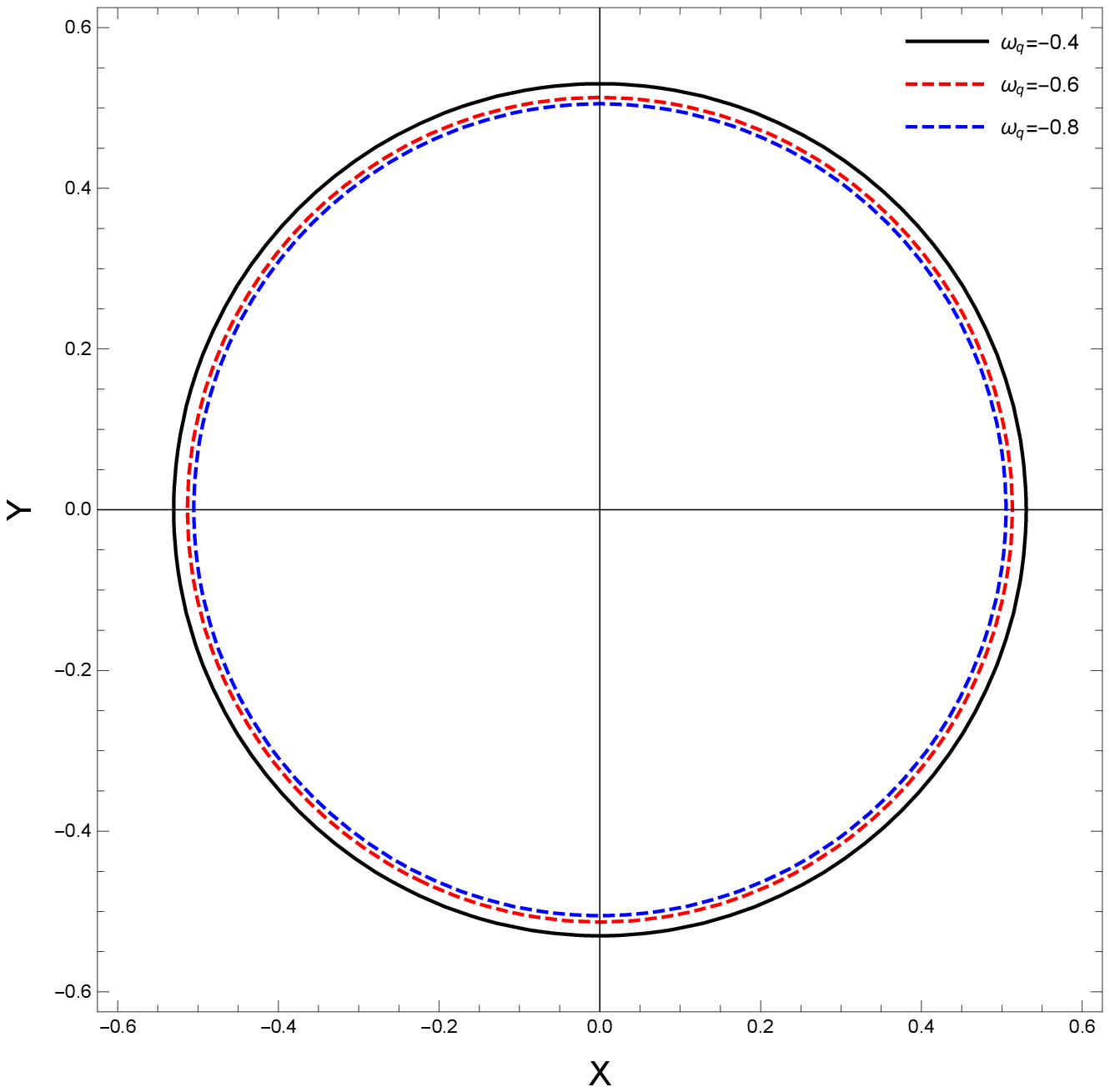}\\
        \subcaption{ $a=0.8$, $Q=0.1$, $L=2$ and $\sigma=0.01$.}\label{fig:Shadowd}
    \end{minipage}\hfill    
\caption{Shadows of quantum-corrected R-N black hole  with quintessence matter for different values of $a$ and $\omega _{q}.$} \label{Shad}
\end{figure}

In literature, it is noted that celestial coordinates, $X$ and $Y$, can be employed to visualize the black hole shadow in a proper form on the observer frame \cite{He, Singh, Kumar}.  Here, we consider the celestial coordinates
\begin{eqnarray}
X&=&\lim_{r_{o}\rightarrow \infty }\left( -r_{o}^{2}\sin \theta _{o}\frac{d\phi }{dr}\right), \\
Y&=&\lim_{r_{o}\rightarrow \infty }\left(
r_{o}^{2}\frac{d\theta }{dr}\right)  \label{cor}
\end{eqnarray}
in which $r_o$ denotes the distance between the observer and the black hole, and $\theta_o$ indicates the angular position of the observer with respect to the black hole's plane. Here, we assume that the observer is far away from the black hole, $\left(r_{o}\rightarrow \infty \right)$. Using the geodesic and the celestial coordinates, we directly relate celestial coordinates to the impact parameters by
\begin{eqnarray}
X&=&-\frac{\xi }{\sin \theta _{o}}, \\
Y&=&\pm \sqrt{\eta -\xi ^{2}\cot
^{2}\theta _{o}}.  \label{cel}
\end{eqnarray}
For an observer in the equatorial plane $\theta _{o}=\pi /2$, Eq. (\ref{cel}) reduces to
\begin{equation}
X^{2}+Y^{2}=\eta +\xi ^{2}.
\end{equation}%
In Fig. \ref{Shad}, we demonstrate shadows of quantum-corrected R-N black hole surrounded by quintessence matter for different quantum correction  and quintessence state parameter values. 

We see that quantum corrections and the quintessence matter field affect the shadow. In particular with a smaller quintessence matter parameter, the shadow becomes smaller for $Q<a$ case.

\newpage
\section{Conclusion}
In this manuscript, we investigate quantum-corrected R-N black hole thermodynamics and shadows in the presence of quintessence matter. We find that the ratio of charge and quantum-correction parameter is very important in the thermal quantities. In particular, if this rate is greater than one, then the mass function has a non-zero value at the lower physical value of the event horizon. The quintessence matter field affects the mass function at higher horizon radii. When we study the modified Hawking temperature we see that it becomes zero at smaller horizon radii when quintessence matter takes smaller values. Moreover, we observe that it presents different characteristic behaviors nearby the physical radius in different cases. When the quantum corrections are considered relatively smaller than the quintessence matter field effect, then the Hawking temperature increases for a small interval and after taking the peak value it decreases. Next, we analyze the heat capacity function, and we notice that in that increasing interval the R-N black hole is stable while in the decreasing temperature interval, the black hole is unstable. It is worth noting that the black hole is always unstable when the quantum correction effects are relatively dominant to the quintessence matter field effects. In both cases, we observe remnant mass values. Therefore, we conclude that the quintessence matter field plays an important role to get a black hole remnant.  In the thermodynamic section, we also derive the equation of state function and present the effects of the considered scenarios with isotherm figures in both cases. { Moreover, after a straightforward algebra  we show that the black hole mimics the Van der Waals fluid.} In the next section, we examine the shadow of the black hole with the Carter approach. By considering a separable solution to the Jacobi action, we obtain the photon orbit equations. According to the particular impact parameters, we derive orbit solutions numerically. By plotting the results, we present the shadows and the effects of the considered quantum corrections and the quintessence matter field.

\section*{Acknowledgments}
{The authors are thankful to the anonymous reviewer for the constructive comments.} This manuscript is supported by the Internal  Project,  [2023/2211],  of  Excellent  Research  of  the  Faculty  of  Science  of Hradec Kr\'alov\'e University.

\section*{Data Availability Statements}
The authors declare that the data supporting the findings of this study are available within the article.


\end{document}